\documentclass[aps,prd,eqsecnum,showpacs,superscriptaddress,amsmath,amssymb,twocolumn,preprintnumbers]{revtex4}
\usepackage{hyperref} 
\usepackage{amsmath,bm,graphicx,slashed}
\usepackage{amssymb}

\newcommand{\nn}{\nonumber} 
\newcommand{\beq}{\begin{equation}}
\newcommand{\eeq}{\end{equation}} 
\newcommand{\beqa}{\begin{eqnarray}} 
\newcommand{\eeqa}{\end{eqnarray}}

\def\bi{\begin{itemize}}
\def\ei{\end{itemize}}
\def\be{\begin{equation}}
 \def\ee{\end{equation}}
 \def\bea{\begin{eqnarray}}
 \def\eea{\end{eqnarray}}
 \def\bean{\begin{eqnarray*}}
 \def\eean{\end{eqnarray*}}
\newcommand{\ie}{{\it i.e.}}

\newcommand{\morder}[1]{{\cal O}\left(#1 \right)}
\newcommand{\eq}[1]{(\ref{#1})}

\newcommand{\ave}[1]{\langle{#1}\rangle}

\newcommand{\Qu}{{\rm Q}}
\newcommand{\Qb}{{\rm\bar Q}}
\newcommand{\QQ}{\Qu\Qb}

\newcommand{\jpsi}{J/\psi}
\newcommand{\xf}{x_{\rm F}}
\newcommand{\pt}{p_{_\perp}}

\newcommand{\lsim}{\lesssim} \newcommand{\gsim}{\gtrsim}

\def\COMMENT#1{}

 \def\esim{\,\mathrel{\rlap{\lower0.2em\hbox{$-$}}\raise0.15em\hbox{\footnotesize $\hskip0.04em\sim$}}\,}
 \def\gsim{\mathrel{\rlap{\lower0.2em\hbox{$\sim$}}\raise0.2em\hbox{$>$}}}
 \def\ksim{\mathrel{\rlap{\lower0.2em\hbox{$\sim$}}\raise0.2em\hbox{$<$}}}

\def\xf{x_{_F}}
\def\sqrtspa{\sqrt{s_{_{pA}}}}
\def\sqrtsha{\sqrt{s_{_{hA}}}}

\begin{document}

%==========================================================
\preprint{LAPTH-019/10}\preprint{CERN-PH-TH/2010-116}
\title{Revisiting scaling properties of medium-induced gluon radiation} 

\author{Fran\c{c}ois Arleo}
\affiliation{ 
LAPTH\footnote{Laboratoire d'Annecy-le-Vieux de Physique Th\'eorique, UMR5108}, Universit\'e de Savoie, CNRS \\ BP 110, 74941 Annecy-le-Vieux cedex, France}
\affiliation{CERN, PH-TH Dept. 1211 Geneva 23, Switzerland}
\author{St\'ephane Peign\'e}
\author{Taklit Sami}
\affiliation{SUBATECH, UMR 6457, Universit\'e de Nantes, Ecole des
Mines de Nantes, IN2P3/CNRS \\ 4 rue Alfred Kastler, 44307 Nantes cedex 3, France}

\date{\today}

\begin{abstract}
\vspace{0.2cm}
Discussing the general case of a hard partonic production process, we show that the notion of parton energy loss is not always sufficient to fully address medium-induced gluon radiation. The broader notion of gluon radiation associated to a hard process has to be used, in particular when initial and final state radiation amplitudes interfere, making the medium-induced radiated energy different from the energy loss of any well-identified parton. Our arguments are first presented in an abelian QED model, and then applied to large-$\xf$ quarkonium hadroproduction. In this case, we show that the medium-induced radiated energy is qualitatively similar (but not identical) to the radiative energy loss of an ``asymptotic massive parton'' undergoing transverse momentum broadening when travelling through the nucleus. In particular, it scales as the incoming parton energy, which suggests to reconsider gluon radiation as a possible explanation of large $\xf$ quarkonium suppression in $p$--A collisions. We expect a similar effect in open heavy-flavour and possibly light-hadron hadroproduction at large $\xf$, depending on the precise definition of the nuclear suppression factor in the latter case. 
\vspace{0.1cm}
\end{abstract}
\maketitle

%==========================================================
\section{Introduction and main results}

High-energy collisions involving atomic nuclei give a unique
opportunity to study parton propagation in nuclear matter. Deep
inelastic scattering (DIS) on nuclei and the Drell-Yan (DY) process in
proton-nucleus collisions are usually considered to be good probes of
parton propagation in cold (confined) nuclear matter, whereas
large-$\pt$ hadron or jet production in heavy-ion collisions should be
sensitive to parton propagation in hot quark-gluon plasma (QGP). It
seems clear that in order to provide a sensible interpretation of the
phenomenon of jet-quenching, observed at RHIC~\cite{Adler:2003qi} 
and recently at the LHC~\cite{Aamodt:2010jdCollaboration:2010bu}, 
and considered as a prominent QGP signal, a good theoretical understanding 
of parton propagation in cold nuclear matter is required.
 
An energetic parton travelling in a large nuclear medium undergoes multiple elastic scatterings, which induce gluon radiation. Usually, the amount of such gluon radiation is referred to as the (radiative) {\it energy loss} of the fast parton, and the notion of parton energy loss has been widely used in phenomenological studies of nuclear effects. For instance, the suppression of production rates in A--A or $p$--A compared to $p$--$p$ collisions (after an adequate normalization) observed for various processes in some kinematical regions (DIS at large $z$, large-$\xf$ DY production in $p$--A, large-$\pt$ hadron production in A--A) has been attributed, at least partly, to parton energy loss~\cite{Accardi:2009qv}. Intuitively, due to parton energy loss, a hard QCD process probes the incoming parton distribution functions (PDF) at higher $x$ where they are suppressed, leading to nuclear suppression. 

However, there exist some hard processes where the medium-induced associated radiation {\it cannot} be strictly identified with the energy loss of a well-defined parton. Such processes are those where a color charge (which may be a composite object) is produced nearly collinearly with one of the incoming partons (in the rest frame of the medium, see section~\ref{sec:frame}), making the gluon emission amplitudes {\it before} and {\it after} the hard production vertex {\it coherent}. Quite intuitively, the presence of interference precludes the identification between associated radiation and radiation off a well-defined parton. The notion of {\it radiated energy} associated to a hard process is thus more general than the notion of parton energy loss. The aim of our study is to illustrate the latter statements, using the specific example of large-$\xf$ quarkonium hadroproduction in high-energy $p$--$p$ and $p$--A collisions. 

Our main result can be summarized as follows. The medium-induced gluon radiation associated to large-$\xf$ quarkonium hadroproduction arises from large gluon formation times, scales as the incoming parton energy
(which at large $\xf$ becomes commensurate with the quarkonium energy), and cannot be identified with the energy loss of a well-defined parton. However, it is qualitatively similar to the Bethe-Heitler energy loss of an ``asymptotic'' massive parton. A similar effect should arise in all processes where a fast color charge is produced nearly collinearly with a fast incoming parton. For instance, the effect is expected in large-$\xf$ open heavy flavour production, but not in Drell-Yan production, where the final energetic particle carries no color charge (see section~\ref{sec:disc} for a more detailed discussion). 

In order to motivate our study, let us first shortly review the notion of parton energy loss. The latter has been extensively studied (see Ref.~\cite{Peigne:2008wu} for a heuristic review). It is of course of crucial importance to know the correct parametric dependence of the mean parton energy loss $\Delta E$, which in general depends on the parton properties (in particular its energy $E$ and mass $M$), those of the medium (its size $L$ and density and/or temperature), and on the coupling $\alpha_s$. But as emphasized in Ref.~\cite{Peigne:2008wu}, the radiative loss of an energetic charged particle also depends on the way the particle is produced. When discussing radiative energy loss, two physical situations have to be distinguished:
\bi
\item[(i)] the energetic charge is suddenly accelerated in the medium, at a given initial time $t=0$ (as in DIS for instance). Since a suddenly accelerated particle radiates even in vacuum, in this case the relevant quantity to address nuclear suppression is not the total energy loss $\Delta E$ suffered by the parton, but the additional, {\it medium-induced} loss $\Delta E_{\rm ind} = \Delta E_{\rm med} - \Delta E_{\rm vac}$ occurring in the medium when compared to vacuum;
\item[(ii)] the energetic charge is an asymptotic particle prepared at $t=-\infty$ and detected at $t=+\infty$, travelling in the medium between $t=0$ and $t=L$. In this situation, the medium-induced loss coincides with the total loss.
\ei

The situation (ii) is more natural in QED, where asymptotic charges do exist. In QCD, the only way to ``see'' a color charge is to resolve it inside a colorless hadron, via some hard partonic subprocess. Even in the absence of a nuclear medium, say in $e$--$p$ or $p$--$p$ collisions, the hard process inevitably produces gluon radiation. Thus, the ``medium-induced prescription'' is required, and at first sight, it is the situation (i) which is generic in QCD. For example, in DIS (DY) production, the medium-induced associated radiation can be identified with the medium-induced energy loss of a quark created (annihilated) at $t=0$, obtained by subtracting the loss associated to the hard process in $e$--$p$ ($p$--$p$) from that in $e$--A ($p$--A) scattering. Large-$\pt$ hadron production around mid-rapidity in A--A collisions is also sensitive to the medium-induced loss of a parton created at $t=0$. (See section~\ref{sec:largept} for a discussion of this case.) 

However, as mentioned in the beginning, for some hard processes the
associated medium-induced radiation does not correspond to the energy
loss of a well-defined parton -- in either situation (i) or (ii). To
illustrate this possibility, let us consider large-$\xf$ quarkonium
hadroproduction, and assume for simplicity that the heavy
quark-antiquark $\QQ$ pair is dominantly produced through the gluon
fusion channel, ${gg} \to \QQ$ \footnote{Our results are however
  independent of the type of the incoming parton, see the discussion
  of Section~\ref{sec:disc}.}. When the longitudinal momentum
$p_{_\parallel}$ of the $\QQ$ pair (in the nucleus rest frame) is
large compared to its transverse mass
\be
M_{\perp}\equiv\sqrt{M^2+\pt^2} \ \, , 
\ee
the $\QQ$ pair is ultra-relativistic and produced nearly collinearly
to the projectile parton. Moreover, at large enough $p_{_\parallel}$,
the $\QQ$ pair remains {\it compact} (of size $\sim 1/M_{\perp}$) and
in a {\it color octet} state for a time $t_{\rm octet} \gg
L$ \footnote{\label{foot4} This holds independently of the quarkonium
  production mechanism. However, we will moreover assume that $t_{\rm
    octet} \gg t_{\rm hard} \gg L$, where $t_{\rm hard}$ is the
  coherence time of the partonic subprocess. Thus, our general
  discussion (see section~\ref{sec:quarkonium}) does not directly apply to the Color Singlet Model (CSM) for quarkonium production, where $t_{\rm octet} \sim t_{\rm hard} \sim E/M^2$. Further comments on the CSM and our assumption about $t_{\rm octet}$ will be made in section~\ref{sec:heavyflav}.}, as a result of time dilation in the nucleus rest frame. Under those conditions, the hard production process looks like gluon scattering at small angle in the target rest frame, except that the outgoing ``gluon'' -- the octet $\QQ$ pair -- is massive. This intuitively explains why the medium-induced radiated energy associated to large-$\xf$ quarkonium hadroproduction is qualitatively similar to the medium-induced energy loss of an ``asymptotic massive parton'' crossing the target nucleus (see section~\ref{sec:quarkonium}). In particular, its scaling as the quarkonium energy might explain the strong nuclear suppression of quarkonium hadroproduction observed at large $\xf$.

\begin{figure}[h]
\centering
\includegraphics[scale=0.3]{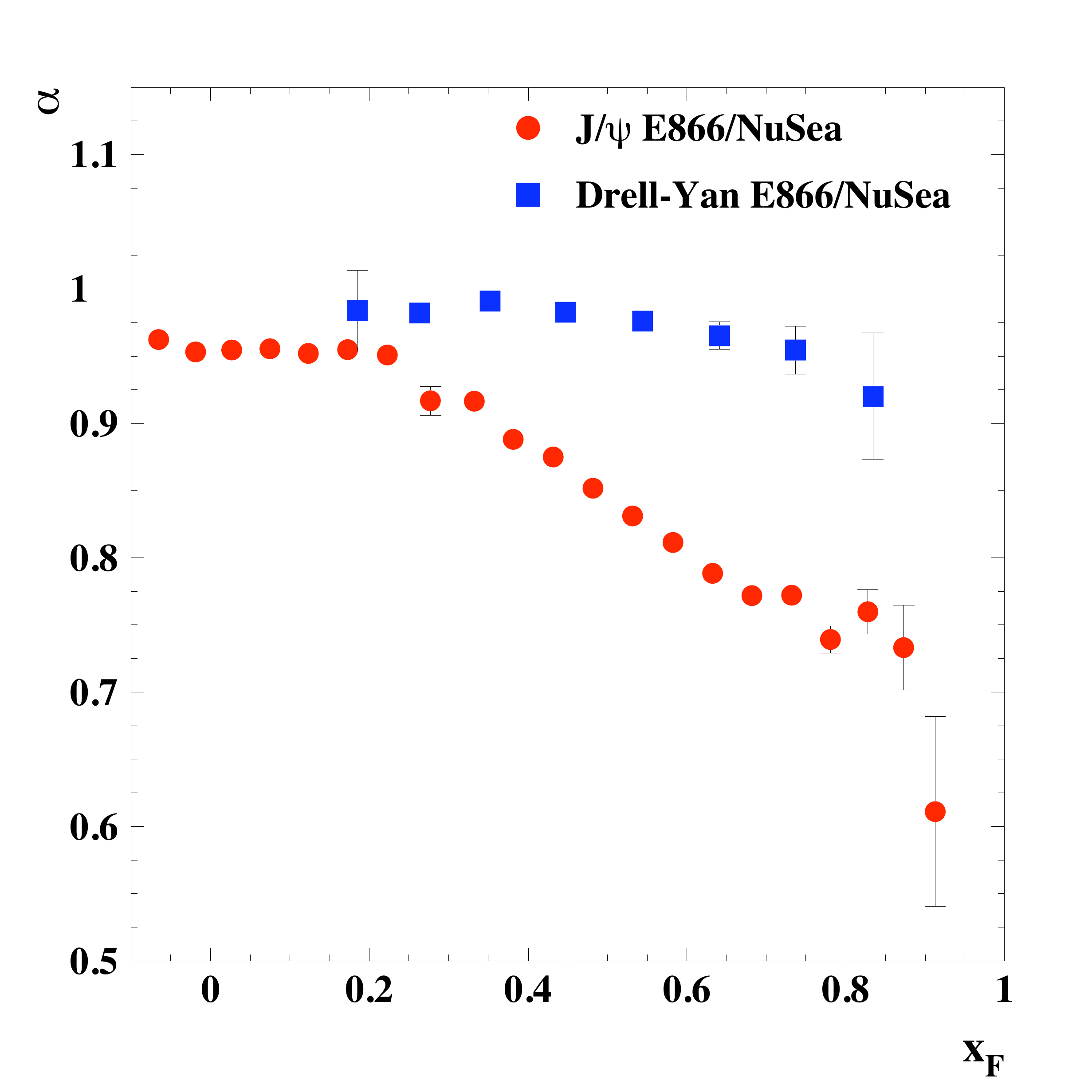}
\caption{Atomic mass number dependence ($\sigma(p$--$A)\propto A^\alpha$) of $\jpsi$ (circles) and Drell-Yan (squares) production ($4<M<8.4$~GeV) measured in $p$--A collisions at $\sqrtspa=38.8$~GeV as a function of $\xf$. E866/NuSea data from~\cite{Leitch:1999ea,Badier:1981ci}.}
\label{fig:data}
\end{figure}

The suppression of $\jpsi$ production in hadron-nucleus collisions at large $\xf$ has been observed by several experiments, at different collision energies $\sqrtsha\simeq20$--40~GeV/nucleon and in various nuclear targets~\cite{Badier:1983dg,Alde:1990wa,Kowitt:1993ns,Leitch:1999ea}. Remarkably, its magnitude is similar to that of open charm measured by E866/NuSea through single muon production~\cite{e866opencharm}, as well as that of light hadrons measured by NA49~\cite{Fischer:2002qp} at SPS and BRAHMS~\cite{Arsene:2003yk} at RHIC; in contrast, it proved much stronger than observed in the DY channel~\cite{Badier:1981ci} as shown in Fig.~\ref{fig:data}, where $\jpsi$ and DY E866/NuSea data are plotted as a function of $\xf$.

%%%%%%%%%%%%%%%%%%%%%%%%%%%%%%%%%%%%%%%%%%%%%%%%%%%%%%%%%%%%%%%%%%%%%%%%%%%%%%
%%%%%%%%%%%%%%%%%%%%%%%%%%%%%%%%%%%%%%%%%%%%%%%%%%%%%%%%%%%%%%%%%%%%%%%%%%%%%%

Various explanations have been proposed to account for these measurements. The comparison of NA3~\cite{Badier:1983dg}, E772/E866~\cite{Alde:1990wa,Leitch:1999ea} as well as PHENIX~\cite{Adler:2005ph} measurements allowed to rule out a scaling of $\jpsi$ suppression in the target parton momentum fraction $x_2$, indicating a violation of factorization in charm hadroproduction~\cite{Hoyer:1990us}. This observation consequently ruled out the nuclear modifications of parton densities in the target as being a dominant effect. On the contrary, the observed scaling in either $\xf$ or $x_1$ supported the qualitative predictions of the intrinsic charm picture in which the soft scattering of higher-Fock states (e.g. $|uudc\bar{c}\rangle$) on the nucleus leads to $A^{2/3}$ scaling ($\alpha=2/3$) at large $\xf$~\cite{Brodsky:1989ex}. Although intrinsic charm might contribute significantly to the observed $\jpsi$ suppression even at rather small $\xf$, its sole effect cannot explain the data given the present constraints on the amount of charm in the proton~\cite{Vogt:1999dw}. The inelastic interaction of the $c \bar c$ pair with nuclear matter seems also disfavoured, since it would require unrealistically large cross sections~\cite{Lourenco:2008sk}; also, $\jpsi$ nuclear absorption models would naturally lead to $x_2$ scaling~\cite{Arleo:1999af} unlike what is observed experimentally.

It has also been suggested by Gavin and Milana (GM) that parton energy loss in nuclear matter might be the dominant effect responsible for both DY and $\jpsi$ suppression~\cite{Gavin:1991qk}. We believe that the present paper should resolve some confusion concerning their explanation. In the GM model, the energy loss is assumed to scale with the incoming parton energy (without real justification), $\Delta E \propto E$, leading to a shift $\Delta x_1 \propto x_1$ in the projectile PDF. This is a key assumption in order to reproduce the qualitative trend of the data. If, on the contrary, energy loss was {\it independent} of the parton energy, the shift in the momentum fraction would not depend on $x_1$, thus leading to a less steep $\jpsi$ suppression as a function of $x_1$, unlike the trend of the data~\cite{Vogt:1999dw}. The stronger $\jpsi$ suppression, compared to that of DY, arises from several effects in GM~\cite{Gavin:1991qk}. Since $\jpsi$ production proceeds mostly via gluon fusion (at not too large $\xf$), the typical energy loss is expected to be larger than for quarks by a factor $N_c/C_F = 9/4$. Another effect comes from the large-$x$ PDF being steeper for gluons than for quarks. Finally, the produced $c \bar c$ pair produced at large $\xf$ travels as a color octet through the nucleus, thus {\it losing energy as a gluon} provided the multiple collisions in the final state do not resolve the $c\bar{c}$ pair. The latter effect appears to be quantitatively crucial as it accounts for more than half of the $\jpsi$ suppression. In summary, in the GM model, $\jpsi$ suppression is due to the initial (quark or gluon) and final (color octet $c \bar c$ pair) in-medium energy losses {\it added incoherently}.

With this in mind, it seems clear that the ``parton energy loss'' implied in the GM model refers to the medium-induced energy loss of a color charge created (or annihilated) at the hard production time $t=0$, \ie, to the situation (i) defined previously. As we will now recall, such parton energy loss does not scale with the parton energy $E$ when $E \to \infty$ (all other scales being fixed), which apparently rules out the Gavin-Milana model. Consider a parton created at some given time $t=0$. As already mentioned, the initial acceleration produces radiation since the initially bare parton progressively builds its proper field, a process which goes along with associated radiation. This happens even when the parton is created in vacuum. In the presence of a nuclear medium, the medium-induced radiation arises due to the rescatterings off target partons shaking off the components of the parton proper field just built. Over the distance $L$, only the components of the proper field which are already formed can be shaken off, hence the medium-induced radiation must satisfy the constraint
\be
t_f \sim \frac{1}{\omega \theta^2} \simeq \frac{\omega}{k_\perp^2} \lsim L \, ,
\label{constraint}
\ee
where $\omega$ and $k_\perp$ are the radiated gluon energy and transverse momentum, $\theta \simeq k_\perp/\omega \ll 1$ the emission angle, and $t_f$ the gluon formation time. The constraint \eq{constraint} yields a bound on parton energy loss, as first derived by Brodsky and Hoyer~\cite{Brodsky:1992nq}, $\Delta E \sim \omega \lsim L \ave{k_\perp^2}$. As noted in Ref.~\cite{Brodsky:1992nq}, the latter bound is a direct consequence of the uncertainty principle, $\Delta L \, \Delta p_z > 1$, implying that some radiation can be released over the length $L$ provided the emission process involves a large enough longitudinal momentum transfer, $p_z > \Delta p_z > 1/ \Delta L > 1/L$. With $p_z \simeq k_\perp^2/(2 \omega)$, one recovers \eq{constraint}. We stress that the Brodsky-Hoyer bound applies to situation (i), \ie, to the medium-induced energy loss of a parton created (or annihilated) in the medium. In this situation, the medium-induced radiation {\it probes} the medium size $L$, as stated by \eq{constraint}. In other words, large formation times $t_f \gg L$ cancel out in the medium-induced loss. Those statements were later confirmed by explicit calculations of the medium-induced radiative loss of a parton created in a medium, showing that $\Delta E \propto L^2 E^0$ when $E \to \infty$~\cite{Baier:1996kr}. 

As legitimately claimed by Brodsky and Hoyer~\cite{Brodsky:1992nq}, the bound on parton energy loss seems to rule out the Gavin-Milana ``energy loss explanation'' of $\jpsi$ nuclear suppression (which uses {\it ad hoc} parton energy losses corresponding to situation (i) and nevertheless scaling as $E$). However, we believe this conclusion relies on a misinterpretation of the physics at work. As we already emphasized, large-$x_F$ quarkonium production is a process where the emission amplitudes off the nearly collinear incoming and outgoing color charges strongly interfere. Thus, the gluon radiation spectrum associated to the hard process cannot be identified with the radiation spectrum off a well-defined parton (or with the incoherent sum of such spectra), contrary to what is assumed in Ref.~\cite{Gavin:1991qk}. 

This is best illustrated by our calculation of section~\ref{sec:largexf}, where large-$\xf$ quarkonium production is modelled by a simple QED process, where an energetic muon of mass $M$ is produced in the hard scattering of an incoming electron of mass $m$, see Fig.~\ref{fig:QEDlargexf}. The (photon) radiation spectrum associated to the hard process in a $p$--$p$ collision is given by (see Fig.~\ref{fig:QEDlargexfrad} and \eq{ppspec-largexf})
\bea
\omega \left. \frac{dI}{d\omega} \right|_{\rm pp} &\simeq& \frac{\alpha}{\pi} \left[ \ln{\left(\frac{1}{\theta_m^2}\right)} + \ln{\left(\frac{1}{\theta_{M}^2}\right)} \right. \nn \\
&& \left. \ \ \ - 2 \ln{\left(\frac{1}{\theta_M^2+ \left. \theta_s^2 \right|_{\rm pp}} \right)}  \right] \, ,
\label{intro-ppspec-largexf}
\eea
where the last term stands for the interference alluded to above, $\left. \theta_s \right|_{\rm pp} = q_\perp /E$ denotes the angle of the final muon with respect to the incoming electron, and $\theta_M \equiv M/E$, $\theta_m \equiv m/E$. The radiation spectrum is similar to the Bethe-Heitler spectrum of an asymptotic charge of mass $M$ undergoing a scattering of angle $\theta_s = \left. \theta_s \right|_{\rm pp}$, see \eq{specasymmuon2}, the latter being approximately obtained by setting $m = M$ in \eq{intro-ppspec-largexf}. Although the associated spectrum \eq{intro-ppspec-largexf} and Bethe-Heitler spectrum are not strictly identical, they share the same property which is essential to our discussion. They both arise from large photon formation times, $t_f \sim 1/(\omega \theta^2) \gg L$, resulting in an integrated spectrum scaling as $E$.

Obviously, the same is true for the spectrum associated to the hard process in a $p$--A collision, obtained from \eq{intro-ppspec-largexf} by the substitution $\left. \theta_s \right|_{\rm pp} \to \left. \theta_s \right|_{\rm pA}$ (see \eq{pAspec-largexf}),
\bea
\omega \left. \frac{dI}{d\omega} \right|_{\rm pA} &\simeq& \frac{\alpha}{\pi} \left[ \ln{\left(\frac{1}{\theta_m^2}\right)} + \ln{\left(\frac{1}{\theta_{M}^2}\right)} \right. \nn \\
&& \left. \ \ \  - 2 \ln{\left(\frac{1}{\theta_M^2+ \left. \theta_s^2 \right|_{\rm pA} }\right)}  \right]  \, .
\label{intro-pAspec-largexf}
\eea
Our main observation is that the {\it medium-induced} spectrum, obtained by subtracting \eq{intro-ppspec-largexf} from \eq{intro-pAspec-largexf}, also arises from large formation times $t_f \gg L$, and moreover solely from the interference terms,
\be
\omega \left. \frac{dI}{d\omega} \right|_{\rm ind} \simeq \frac{2 \alpha}{\pi}  \, \ln{\left(1+\frac{\Delta \theta_s^2}{\theta_M^2+ \left. \theta_s^2 \right|_{\rm pp} }\right)}  \, ,
\label{intro-indspeclargexf3}
\ee
where $\Delta \theta_s^2 = \left. \theta_s^2 \right|_{\rm pA} - \left. \theta_s^2 \right|_{\rm pp}$. As a consequence, the medium-induced radiated energy associated to the large-$\xf$ production process scales as 
the energy $E$ of the incoming charge, similarly to Bethe-Heitler radiation. This conclusion trivially generalizes to the QCD case of quarkonium hadroproduction (see section~\ref{sec:quarkonium}). Thus, the Gavin-Milana assumption of an ``energy loss'' scaling as $E$ turns out to be valid for quarkonium production \footnote{The GM assumption is however incorrect in DY production at large $\xf$, where no energetic and quasi-collinear color charge is produced in the final state. The radiated energy associated to the Drell-Yan process can thus be identified with the energy loss of the incoming and suddenly decelerated quark, see section~\ref{sec:dy}.}, provided this ``energy loss'' is correctly interpreted as the radiated energy associated to the hard process, and not as the energy loss of independent incoming and outgoing color charges. The medium-induced radiated energy is parametrically similar to the Bethe-Heitler energy loss of a charge of mass $M$ created in the far past, in particular it arises from large formation times, and the Brodsky-Hoyer bound does not apply in this case. 

One might wonder why large formation times $t_f \gg L$ do not cancel out in the medium-induced radiation, as in the case of a parton created in a medium (see \eq{constraint}). Radiation associated with $t_f \gg L$ does not {\it resolve} the medium, but this does not always imply that the radiation is independent of the medium size and properties. For a parton created in the medium, it does. Indeed, in this case, radiation with $t_f \gg L$ solely arises from late emission off the {\it final} charged particle, and is the same with or without medium. This is why formation times $t_f \gg L$ cancel out in the medium-induced energy loss of a parton created in the medium. In the case of large-$\xf$ quarkonium hadroproduction (and similarly, in the case of Bethe-Heitler radiation off an asymptotic charge), although radiation with $t_f \gg L$ does not resolve the medium and sees it as a pointlike object, it however depends on the medium size via the transverse momentum broadening $\Delta q_\perp^2 \propto L$. Thus, the contribution with $t_f \gg L$ does {\it not} cancel in the medium-induced radiation associated to large-$\xf$ quarkonium hadroproduction, contrary to what is assumed in Ref.~\cite{Brodsky:1992nq}. 

The effect discussed here differs from that studied in Refs.~\cite{Kopeliovich:2005ym}. In those studies, the nuclear suppression at large $\xf$ is explained from simple energy conservation arguments. In the limit $\xf \to 1$, the production of additional particles is forbidden, and each rescattering of the (initial and/or final) energetic parton is associated to a Sudakov factor $\sim(1-\xf)$. Since there are more rescatterings in a nucleus than in a proton target, this naturally leads to nuclear suppression at large $\xf$. This effect appears for all processes, including those (like DIS and DY production) where no energetic color charge is present in the initial or final state of the hard subprocess. It is argued in Refs.~\cite{Kopeliovich:2005ym} that the effect acts similarly to an {\it effective} parton energy loss scaling as the parton energy, independently of the process under consideration. This is clearly different from the actual parton energy loss (and more generally associated radiation) studied in the present paper. Although energy conservation is obviously more and more important when $\xf$ approaches unity and has to be implemented in any realistic phenomenological model, we expect our effect to play a crucial role even far away from the edge of phase-space and therefore independently of the constraints from energy conservation. The implications of our results on phenomenology will be studied in a future work. 

The paper is organized as follows. In section~\ref{sec:hardproc} we discuss the radiated energy associated to a hard process, in a simple QED model. In section~\ref{sec:largept} we consider the case of large angle scattering, where the radiated energy can be identified with the energy loss of a given charge, as commonly assumed. In section~\ref{sec:largexf}, we present a simple, small-angle scattering process where this identification is not possible due to the presence of interference between initial and final state radiation. This situation is generalized in section~\ref{sec:quarkonium} to the QCD case of large-$\xf$ quarkonium hadroproduction, where we obtain the non-abelian analog of \eq{intro-indspeclargexf3} for the medium-induced radiation spectrum. Some more detailed questions, such as the dependence of our results on the precise quarkonium production mechanism, as well as the comparison with other processes (e.g. open heavy flavour and light hadron production) are addressed in section~\ref{sec:disc}.

%==========================================================
\section{Radiation spectrum associated to a hard process}
\label{sec:hardproc}

\subsection{Preliminary considerations}
\label{sec:frame}

\begin{figure}[h]
\centering
\includegraphics[scale=0.6]{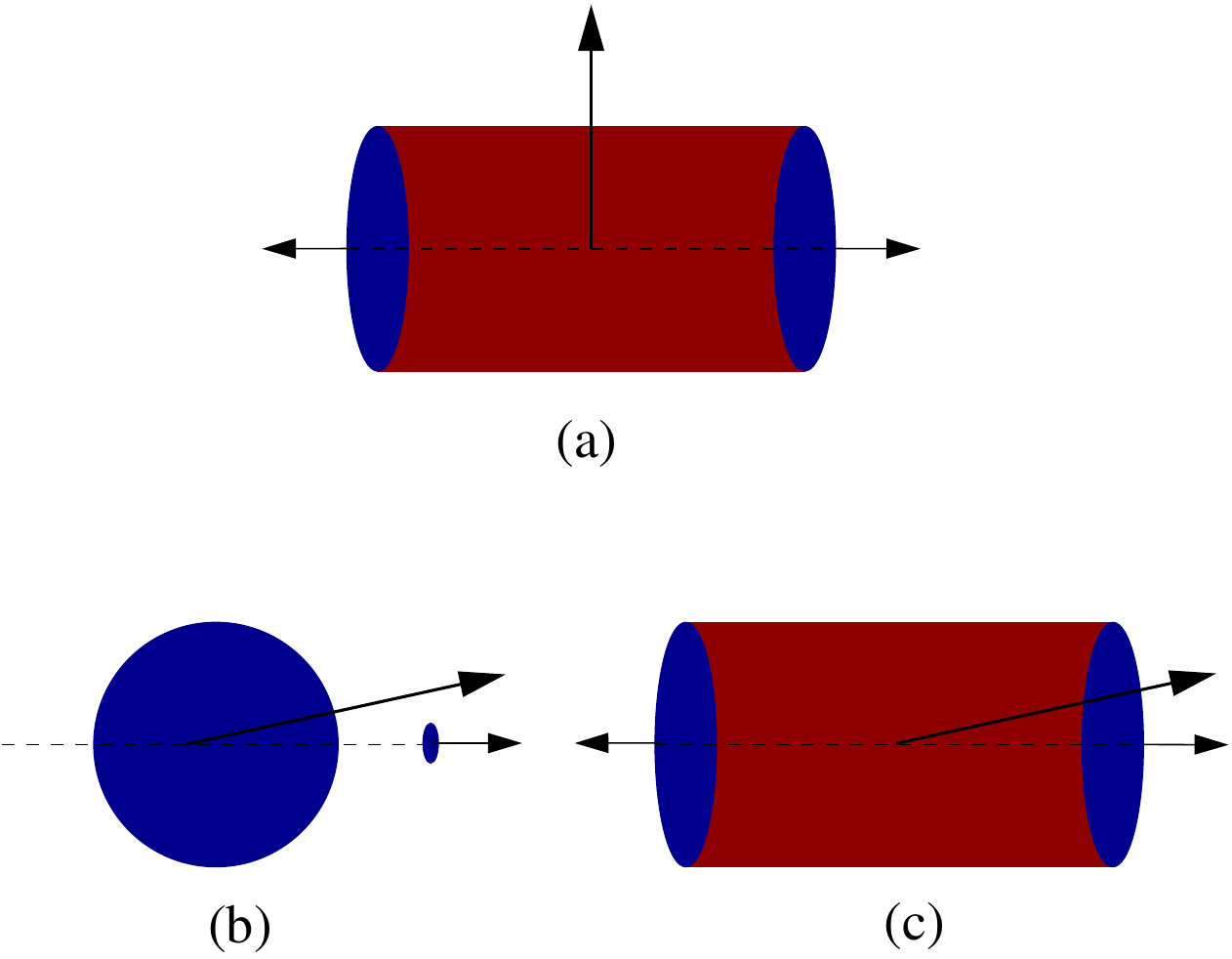}
\caption{{\it Large angle:} (a) Particle produced at mid-rapidity in A--A collisions. {\it Small angle:} (b) particle produced in $p$--A collisions, (c) particle produced at large rapidity in A--A collisions.}
\label{fig:sketch}
\end{figure}

In the following discussion we shall consider the medium-induced gluon radiation spectrum, $\omega dI/d\omega$, arising from the transverse momentum broadening of partons propagating through a QCD medium. The natural Lorentz frame to study this problem is therefore the frame in which the medium is {\it static}, denoted by (S). This frame is different whether parton propagation occurs in hot quark-gluon plasma (A--A collisions) or cold nuclear matter (say, $p$--A collisions). In the former case, the frame (S) is the center-of-mass frame of the heavy-ion reaction, neglecting for simplicity the longitudinal and transverse expansion of the plasma. The frame (S) of the latter is the target nucleus rest frame, boosted by a Lorentz factor $\gamma = \sqrt{s}/(2m_p)\gg1$ with respect to the c.m. frame of the $p$--A collision.
In the static frame (S), two kinematical situations can be considered:
\begin{itemize}
\item[(i)] {\it large angle scattering}, discussed in~\ref{sec:largept}. It corresponds typically to the production of large-$\pt$ particles produced around mid-rapidity and propagating in QGP, as illustrated in Fig.~\ref{fig:sketch}a;
\item[(ii)] {\it small angle scattering}, discussed in~\ref{sec:largexf}, which occurs in two distinct cases, namely parton propagation in cold nuclear matter (since all momenta are mostly longitudinal in the target nucleus rest frame from the large boost), see Fig.~\ref{fig:sketch}b, as well as large-rapidity particle propagation in QGP, Fig.~\ref{fig:sketch}c.
\end{itemize}

%%%%%%%%%%%%%%%%%%%%%%%%%%%%%%%%%%%%%%%%%%%%%%%%%%%%%%%%%%%%%%%%%%%%%%%%%%%%%%
%%%%%%%%%%%%%%%%%%%%%%%%%%%%%%%%%%%%%%%%%%%%%%%%%%%%%%%%%%%%%%%%%%%%%%%%%%%%%%

To simplify the discussion in sections~\ref{sec:largept} and~\ref{sec:largexf}, we consider QED processes. The general conclusions drawn also hold in QCD. 

\subsection{Large angle scattering}
\label{sec:largept}

\begin{figure}[h]
\centering
\includegraphics[scale=0.9]{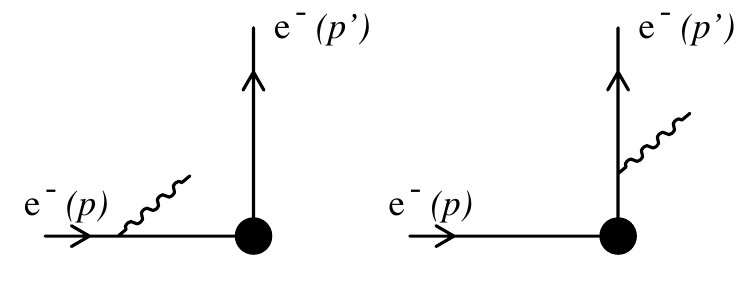}
\caption{Model for large angle scattering in QED in ``$p$--$p$'' collisions. The blob represents the hard process, the photon is radiated by the incoming (left) and outgoing (right) electron.}
\label{fig:pplargept-radamp}
\end{figure}

Here we discuss large angle scattering in the frame (S), and first consider ``$p$--$p$'' collisions. The hard partonic process is modelled by an incoming electron, of mass $m$ and energy $E \gg m$, scattering at large angle ($90^\circ$ in the frame (S)). The photon radiation spectrum associated to the hard process is obtained by calculating the photon emission amplitudes off the incoming and outgoing electron lines represented in Fig~\ref{fig:pplargept-radamp}. 

%%%%%%%%%%%%%%%%%%%%%%%%%%%%%%%%%%%%%%%%%%%%%%%%%%%%%%%%%%%%%%%%%%%%%%%%%%%%%%
%%%%%%%%%%%%%%%%%%%%%%%%%%%%%%%%%%%%%%%%%%%%%%%%%%%%%%%%%%%%%%%%%%%%%%%%%%%%%%

In the limit of soft photon energy $\omega \ll E$, the photon emission vertices factorize from the hard amplitude, and one can write the photon radiation intensity as 
\be
dI = \sum_{i=1,2} e^2 \left| \frac{p \cdot \varepsilon_i}{p\cdot k} - 
\frac{p' \cdot \varepsilon_i}{p'\cdot k}\right|^2 
\, \frac{d^3\vec{k}}{(2\pi)^3 2\omega} \, , 
\label{photonintensity} 
\ee
where $p=(E,\vec{p})$ and $p'=(E',\vec{p}{\, '})$ are the four-momenta of the incoming and outgoing electrons, $k=(\omega,\vec{k})$ is the photon four-momentum, $\varepsilon_i=(0,\vec{\varepsilon}_i)$ are two transverse photon polarization vectors, and $e$ is the QED coupling. The angular integral of \eq{photonintensity} can be performed exactly~\cite{Weinberg:1965nx}, yielding the soft photon energy spectrum 
\be
\omega \frac{dI}{d\omega} = \frac{2 \alpha}{\pi} \left[ {\cal R}(v) + {\cal R}(v') + {\cal I}(v, v', \vec{v} \cdot \vec{v}{\, '}) \right] \, , 
\label{softspec}
\ee
where ${\cal R}(v)$ (respectively ${\cal R}(v')$) corresponds to the square of the emission amplitude off the initial (final) electron line, and ${\cal I}(v, v', \vec{v} \cdot \vec{v}{\, '}) $ stands for the interference term. It is worth recalling that these three terms only depend on the initial and final electron velocities $\vec{v} \equiv \vec{p}/E$ and  $\vec{v}{\, '} \equiv \vec{p}{\, '}/E'$. We have
\bea
{\cal R}(v) &=& \frac{1}{2v} \ln{\left( \frac{1+v}{1-v}\right)} -1  \, ,
\label{Rofv} \\
{\cal I}(v, v', \vec{v} \cdot \vec{v}{\, '}) &=& {\cal R}(\beta)- {\cal R}(v) - {\cal R}(v') \, ,
\label{int} 
\eea
where $\beta$ is the relative velocity of the final electron in the rest frame of the initial one, 
\be
\beta(v, v', \vec{v} \cdot \vec{v}{\, '}) = \sqrt{1-\frac{(1-v^2)(1-v'^2)}{(1-\vec{v} \cdot\vec{v}{\, '})^2}} \, .
\label{beta}
\ee
Hence, the energy spectrum reads
\be
\omega \frac{dI}{d\omega} = \frac{2 \alpha}{\pi} {\cal R}(\beta) \, .
\label{softspec1}
\ee
This has the following simple interpretation. In the rest frame of the initial electron, the final electron is suddenly accelerated (to the velocity $\beta$), and the spectrum must be fully given by the square of the emission amplitude off the final electron line. 

We focus on the ultrarelativistic limit where $v$, $v' \to 1$. In this limit, 
\be
{\cal R}(v)  \simeq \frac{1}{2} \ln{\left( \frac{1}{1-v} \right)} \to \infty \, ,
\label{col-log}
\ee
showing that the squares of the initial and final emission amplitudes suffer from a logarithmic collinear singularity. This corresponds to DGLAP radiation~\cite{Dokshitzer:1977sg}, which in QCD is resummed in the initial (final) parton distribution (fragmentation) function. When $v$, $v' \to 1$ and the angle between $\vec{v}$ and $\vec{v}{\, '}$ is fixed, we have $\beta \to 1$, and the interference term reads
\be
{\cal I}(v, v', \vec{v} \cdot \vec{v}{\, '}) \mathop{\simeq}_{\beta \to 1} 1 + \ln{\left( \frac{1-\vec{v} \cdot \vec{v}{\, '}}{2} \right)} \, .
\label{int-betatoone}
\ee
This shows no logarithmic enhancement when $\vec{v}$ and $\vec{v}{\, '}$ are quasi-orthogonal ($\vec{v} \cdot\vec{v}{\, '} \ll 1$).

Thus, in the kinematical situation of Fig.~\ref{fig:pplargept-radamp}, the associated radiation spectrum is dominated by the squares of the initial and final emission vertices, due to a logarithmic enhancement of collinear emission. At ultrarelativistic energies, the radiation spectrum associated to large angle scattering is given, in the leading logarithmic approximation, by the sum of emission spectra of well-identified ``partons'', namely those of an incoming (suddenly decelerated) and outgoing (suddenly accelerated) electron. We stress that the dominance of collinear logarithms implies the dominance of large photon formation times $t_f$ in the radiation spectrum. For instance, the logarithm in \eq{col-log} arises from the angular domain $\theta_m^2 \ll \theta^2 \ll 1$, where $\theta$ is the photon emission angle with respect to the initial electron, and $\theta_m^2 = m^2/E^2 = 1-v^2$. The photon formation time $t_f \sim 1/(\omega \theta^2)$ is thus very large in the ultrarelativistic limit. 

The identification between radiation associated to large angle scattering and parton radiation trivially extends to the case of heavy-ion collisions. Consider some (abelian) QGP of finite size $L$ produced in a central A--A collision, shortly after the time of the hard ``partonic'' process of Fig.~\ref{fig:pplargept-radamp}. In the hot environment, the final electron undergoes {\it soft} rescatterings, which will modify the radiation spectrum associated to the hard process. Let us assume that large photon formation times $t_f \gg L$ dominate in the radiation spectrum. Then, photon radiation does not probe the medium, and the radiation amplitude is given by the sum of the emissions off the initial and final electron lines, as in the absence of a medium. The in-medium rescatterings affect dominantly the {\it direction} of the final electron velocity, and the radiation spectrum in the presence of a medium is given by
\bea
\left. \omega \frac{dI}{d\omega}\right|_{\rm med} &=& \frac{2 \alpha}{\pi} \left[ {\cal R}(v) + {\cal R}(v') + {\cal I}(v, v', \vec{v} \cdot \vec{v}{\, '} + \delta \,\vec{v} \cdot \vec{v}{\, '}) \right] \nn \\
&\simeq& \frac{2 \alpha}{\pi} \left[ {\cal R}(v) + {\cal R}(v') \right] \, ,
\label{softspecmed}
\eea
where $\delta \,\vec{v} \cdot \vec{v}{\, '}$ arises from the in-medium modification of the final electron direction. For $\vec{v} \cdot \vec{v}{\, '}= 0$ and $\delta \,\vec{v} \cdot \vec{v}{\, '} \ll 1$, the interference term has no logarithmic enhancement, and arises from small formation times. Hence, it must be dropped in the first line of \eq{softspecmed}, which was obtained assuming $t_f \gg L$. As in the vacuum case, the spectrum is dominated by the collinear logarithms of the squared terms, and thus by large photon formation times $t_f \gg L$. 

In the phenomenological analyses of nuclear effects, production rates in A--A collisions are normalized by the similar rates in $p$--$p$ collisions. Then, jet-quenching does not depend on the total radiation associated to the hard process, but rather on the {\it additional} radiation occurring with a medium, when compared to the ``vacuum'', $p$--$p$ case. In the following we will focus on the so-called {\it medium-induced} radiation spectrum, obtained by subtracting the vacuum contribution from the in-medium spectrum. Subtracting \eq{softspec} from \eq{softspecmed}, the dominant (collinear) terms cancel out, removing the contribution of large formation times $t_f \gg L$ in the induced spectrum. Hence, the {\it medium-induced} radiation associated to large angle scattering must originate from photons with li\-mi\-ted formation time, $t_f \lsim L$. Since the expression \eq{softspecmed} was obtained assuming that large formation times do\-mi\-na\-te, it is not adequate to derive the medium-induced spectrum. More work is needed to correctly derive the latter, as we now recall. 

\begin{figure}[h]
\centering
\includegraphics[scale=0.8]{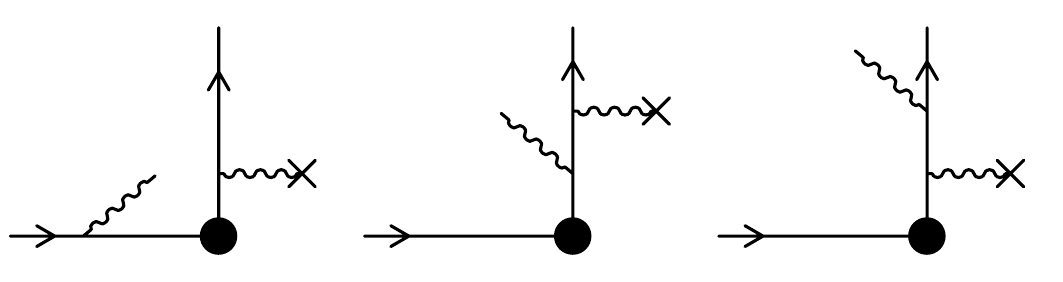}
\caption{Photon emission diagrams in ``A--A'' collisions, where the outgoing electron rescatters in the medium.}
\label{fig:AAradamp}
\end{figure}

Photons with $t_f \lsim L$ can probe the medium size $L$, and the radiation amplitude off the scattered electron is not simply given by the coupling to the external lines. In general the induced spectrum depends on the details of the electron rescatterings in the medium. For the purpose of the present discussion it is sufficient to consider the case of a small medium of size $L \ll \lambda$, with $\lambda$ the electron mean free path in the medium, so that the scattered electron undergoes at most one elastic scattering. The radiation amplitude induced by such a scattering is given by three diagrams represented in Fig.~\ref{fig:AAradamp}. 

%%%%%%%%%%%%%%%%%%%%%%%%%%%%%%%%%%%%%%%%%%%%%%%%%%%%%%%%%%%%%%%%%%%%%%%%%%%%%%
%%%%%%%%%%%%%%%%%%%%%%%%%%%%%%%%%%%%%%%%%%%%%%%%%%%%%%%%%%%%%%%%%%%%%%%%%%%%%

\begin{figure}[h]
\centering
\includegraphics[scale=1]{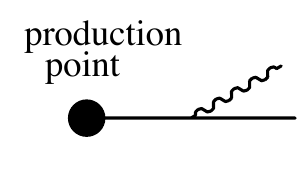}
\caption{An electron produced in the vacuum at an initial time $t=0$ by a hard process radiates a photon.}
\label{fig:pprad}
\end{figure}

\begin{figure}[h]
\centering
\includegraphics[scale=1]{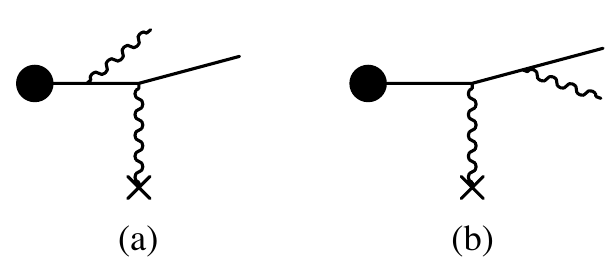}
\caption{Photon radiated by an electron produced at $t=0$ in the medium.}
\label{fig:AArad}
\end{figure}

The diagram where the photon is emitted from the {\it internal} electron line (\ie, between the hard scattering and the soft in-medium rescattering), is negligible when $t_f \gg L$ and thus in the total spectrum \eq{softspecmed}, but becomes important when $t_f \lsim L$. Squaring the sum of the amplitude of Fig.~\ref{fig:AAradamp}, and subtracting the vacuum contribution (see Fig.~\ref{fig:pplargept-radamp}), we find that the square of the emission amplitude off the incoming electron cancels out, and that the (induced) interference between the emissions before and after the hard production vertex is negligible. Thus, in order to derive the medium-induced radiation spectrum associated to large angle scattering, we can discard the incoming electron, and simply consider the radiation amplitudes of Fig.~\ref{fig:pprad} (for $p$--$p$ collisions) and of Fig.~\ref{fig:AArad} (for A--A collisions), where the blob represents the hard scattering. The medium-induced spectrum is that of an electron ``created'' at the hard scattering time $t=0$, and propagating through the medium. 

%%%%%%%%%%%%%%%%%%%%%%%%%%%%%%%%%%%%%%%%%%%%%%%%%%%%%%%%%%%%%%%%%%%%%%%%%%%%%%
%%%%%%%%%%%%%%%%%%%%%%%%%%%%%%%%%%%%%%%%%%%%%%%%%%%%%%%%%%%%%%%%%%%%%%%%%%%%%%
%%%%%%%%%%%%%%%%%%%%%%%%%%%%%%%%%%%%%%%%%%%%%%%%%%%%%%%%%%%%%%%%%%%%%%%%%%%%%

In order to display the parametric dependence of the medium-induced spectrum, let us review the derivation of Ref.~\cite{Peigne:2008wu} (for $L \ll \lambda$). In vacuum, the radiation spectrum is obtained by squaring the diagram of Fig.~\ref{fig:pprad}. Neglecting the electron mass and working in the small angle approximation, the spectrum reads 
\be
\omega \left. \frac{dI}{d\omega} \right|_{\rm vac} = \frac{\alpha}{\pi^2} \int d^2\vec{\theta} \, \vec{J}_{\rm vac}^{\ 2} \ \ \ ; \ \ \ \vec{J}_{\rm vac} \equiv \frac{\vec{\theta}}{\theta^2} \, ,
\label{ppspec}
\ee
where the vector $\vec{\theta} \equiv \vec{k}_{\perp}/\omega$ denotes the photon ``angle'' with respect to the final electron. The radiation spectrum induced by a single in-medium rescattering is obtained from \eq{ppspec} by replacing $\vec{J}_{\rm vac} \to \vec{J}_{\rm med}$, where the in-medium emission current reads~\cite{Peigne:2008wu}
\be
\vec{J}_{\rm med} = \frac{\vec{\theta}'}{\theta'^2} - \frac{\vec{\theta}}{\theta^2}
\left[ 1-  e^{- i\omega L_0 \theta^2/2 } \right] \, .
\label{Jmed}
\ee
Here $L_0$ is the distance travelled by the electron between its production point and the scattering. When the electron is produced in the medium, we have $L_0 \sim L$. The term $\propto \vec{\theta}$ corresponds to the graph in Fig.~\ref{fig:AArad}a and the term $\propto \vec{\theta'}$ to the graph in Fig.~\ref{fig:AArad}b. $\vec{\theta}'$ is the angle between the photon and the electron direction {\it after} the rescattering, $\vec{\theta}' = \vec{\theta} - \vec{\theta}_s$, where $\vec{\theta}_s \equiv \vec{q}_\perp/E$ is the electron scattering angle ($\theta_s \ll 1$) and $\vec{q}_\perp$ the transverse momentum exchange in the elastic scattering.

The medium-induced spectrum is obtained by subtracting the vacuum from the in-medium spectrum, and multiplying by the single scattering probability $\sim L/\lambda$, 
\be
\omega \left. \frac{dI}{d\omega} \right|_{\rm ind} \sim \frac{L}{\lambda} \cdot \frac{\alpha}{\pi^2} \int d^2\vec{\theta} \, \left( \vec{J}_{\rm med}^{\ 2}   - \vec{J}_{\rm vac}^{\ 2} \right) \, .
\label{inducedspec1}
\ee
Using the expressions of $\vec{J}_{\rm vac}$ and $\vec{J}_{\rm med}$, and the identity
\be
\int d^2\vec{\theta} \, \left[ \left| \frac{\vec{\theta}'}{\theta'^2} \right|^2 - \left| \frac{\vec{\theta}}{\theta^2} \right|^2 \right] = 0 \, ,
\label{identity}
\ee
the induced spectrum can be expressed as 
\be
\omega \left. \frac{dI}{d\omega} \right|_{\rm ind} \sim \frac{L}{\lambda} \cdot \frac{2 \alpha}{\pi^2} \int d^2\vec{\theta} \, \frac{\vec{\theta}}{\theta^2} \cdot \left( \frac{\vec{\theta}}{\theta^2} - \frac{\vec{\theta}'}{\theta'^2}  \right) \left[ 1-\cos{\frac{\omega L_0 \theta^2}{2}} \right] \, .
\label{inducedspec2}
\ee 

The identity \eq{identity} states that the radiation (integrated over angles) occurring {\it after} the electron scattering (\ie, the square of Fig.~\ref{fig:AArad}b) is identical to the vacuum DGLAP radiation. (This is the small angle expression of the cancellation of the ${\cal R}(v')$ term between \eq{softspecmed} and \eq{softspec}.) This is because the radiation after the soft rescattering depends only on the photon angle $\vec{\theta}' = \vec{\theta} - \vec{\theta}_s$ with respect to the final electron. In the angular integration, the shift $\vec{\theta} \to \vec{\theta} + \vec{\theta}_s$ removes all dependence on the scattering, and thus on the size or properties of the medium. Hence, there is an exact cancellation when removing the vacuum part to obtain the medium-induced spectrum, corresponding to the cancellation of large formation times $t_f \gg L$, and the induced spectrum is thus dominated by $t_f \lsim L$. This is a crucial point, which is at the basis of the drastic difference between the present situation and the small angle scattering case studied in section~\ref{sec:largexf}. 

To explicitly verify in the present situation that the medium-induced energy spectrum is indeed dominated by $t_f \lsim L$, let us simplify \eq{inducedspec2} by averaging over $\vec{\theta}_s$ as in Ref.~\cite{Peigne:2008wu}. First, we average over azimuthal directions of $\vec{\theta}_s$ using
\be
\label{azimint}
\int \frac{d\phi}{2\pi} 
\left( \frac{\vec{\theta}}{\theta^2} - \frac{\vec{\theta} - \vec{\theta}_s }{(\vec{\theta} - \vec{\theta}_s)^2} \right) 
= \frac{\vec{\theta}}{\theta^2} \, \Theta(\theta^2_s - \theta^2)  \, .
\ee
We then average over $\theta^2_s$ using the (screened) Coulomb scattering probability distribution
\be
\label{Pthets}
P(\theta_s^2) \ =\ \frac {\mu^2/E^2}{(\theta^2_s + \mu^2/E^2)^2} \ , 
\ee
where $\mu$ is the typical value of the transverse momentum exchange $q_\perp$ in Coulomb scattering. We obtain
\be
\omega \left. \frac{dI}{d\omega} \right|_{\rm ind} \sim \frac{L}{\lambda} \cdot \frac{2\alpha}{\pi} \cdot \frac{\mu^2}{E^2} 
\int_0^\infty d\theta^2 \, \frac{1- \cos(\omega L_0 \theta^2/2)}{\theta^2(\theta^2 + \mu^2/E^2)}  \, .
\label{inducedspec3}
\ee
At high energy we have $\mu^2/E^2 \ll 1/(\omega L_0)$, and the angular integral in \eq{inducedspec3} is saturated by $\theta^2 \sim 1/(\omega L_0)$, \ie, by formation times 
\be
t_f \sim \frac{1}{\omega \theta^2}  \sim L_0 \sim L \, .
\ee
The medium-induced energy spectrum reads
\be
\omega \left. \frac{dI}{d\omega} \right|_{\rm ind}  \sim \alpha \cdot \frac{\omega}{E^2} \cdot \frac{L^2 \mu^2}{\lambda} \, ,
\label{photonspec} 
\ee
and integrating this spectrum up to $\omega \sim E$ yields the medium-induced electron energy loss 
\be
\Delta E_{\rm ind}  \sim \alpha \cdot \frac{\mu^2}{\lambda} \cdot L^2 \, .
\label{t=0electronloss}
\ee
We see that the medium-induced energy loss is energy independent and much smaller than the ``DGLAP energy loss'' (obtained by integrating \eq{ppspec}) which is proportional to $E$. This is a direct consequence of the constraint $t_f \lsim L$ for medium-induced radiation. We stress that the medium-induced spectrum is well-defined in the $m \to 0$ limit, and thus collinear safe. This is because the contributions to the vacuum and in-medium spectra arising from small angles $\theta^2 \sim \theta_m^2 \to 0$, \ie, from large formation times, solely arise from the DGLAP terms which cancel in the induced spectrum. 

The main lesson of this section can be summarized as follows: {\it The medium-induced spectrum associated to large angle scattering can be identified to that of a charged particle created at $t=0$. Large formation times cancel out in the medium-induced spectrum, leaving only $t_f \lsim L$, and resulting in an energy-independent and collinear safe medium-induced energy loss.} 

\subsection{Small angle scattering}
\label{sec:largexf}

We now discuss small angle scattering processes in the frame (S). In the following, the invariant mass $M_X$ of the partonic final state sets the hardness of the process. We work in the target rest frame, where small angle scattering is modelled as an incoming energetic charge of momentum $p = (E,\vec{0}_\perp, p_z)$ scattering off the target with limited momentum exchange $q$. Creating the invariant mass $M_X$ requires some non-zero longitudinal momentum transfer $q_z$, 
\be
(p+q)^2 = M_X^2 \Rightarrow |q_z| \sim \frac{M_X^2}{E} \, ,
\ee
which using the uncertainty principle is related to the coherence time $t_{\rm hard}$ of the hard partonic process,
\be
t_{\rm hard} \sim \frac{1}{|q_z|} \sim \frac{E}{M_X^2} \, .
\label{lhard}
\ee

\begin{figure}[h]
\centering
\includegraphics[scale=1.2]{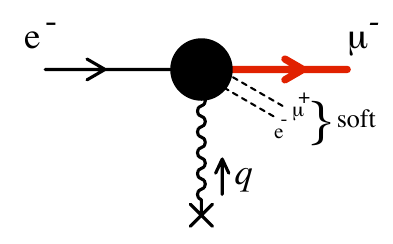}
\caption{QED model for the production of a massive particle (muon) at small angle. The outgoing electron and anti-muon are soft and do not participate to the hard scattering dynamics.}
\label{fig:QEDlargexf}
\end{figure}

We want to study the radiation spectrum associated to a generic small angle scattering process, and see how the spectrum is modified when going from ``$p$--$p$'' to ``$p$--A'' collisions, within a simple QED model. Our arguments are quite general and also apply to non-abelian radiation (despite slight parametric differences), as we will explicitly see in section~\ref{sec:quarkonium} in the QCD case of quarkonium production. Our QED model for small angle scattering is depicted in Fig.~\ref{fig:QEDlargexf}. 

%%%%%%%%%%%%%%%%%%%%%%%%%%%%%%%%%%%%%%%%%%%%%%%%%%%%%%%%%%%%%%%%%%%%%%%%%%%%%%
%%%%%%%%%%%%%%%%%%%%%%%%%%%%%%%%%%%%%%%%%%%%%%%%%%%%%%%%%%%%%%%%%%%%%%%%%%%%%

An electron of mass $m$ and energy $E \gg m$ undergoes a hard scattering, from which a muon of mass $M \gg m$ and energy $E' \simeq E \gg M$ emerges. The particles produced in the final state in conjunction with the energetic muon (which are actually required here from lepton number conservation) are soft. Radiation off those soft particles (for instance, the outgoing $\mu^+$ and $e^-$ in Fig.~\ref{fig:QEDlargexf}) can be disregarded, and only the incoming and outgoing {\it energetic charges} are relevant for our purpose. 

\begin{figure}[h]
\centering
\includegraphics[scale=1.2]{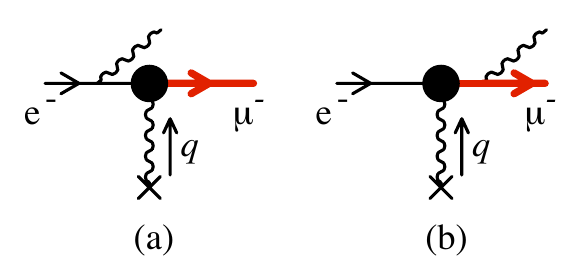}
\caption{Photon emission by the incoming electron and outgoing muon, in the QED model for small angle scattering.}
\label{fig:QEDlargexfrad}
\end{figure}

The radiation spectrum associated to the process of Fig.~\ref{fig:QEDlargexf} is easily obtained by assuming that it arises from photon formation times satisfying $t_f \gg t_{\rm hard}$, which can be checked a posteriori \footnote{The precise value of $t_{\rm hard}$, and thus of the final state invariant mass $M_X$ is irrelevant to our considerations, as long as $t_f \gg t_{\rm hard}$. In Fig.~\ref{fig:QEDlargexf}, $M_X^2 \sim M E$, implying $t_{\rm hard} \sim 1/M$. In quarkonium production, $M_X^2 \sim M^2$, with $M$ the quarkonium mass, implying $t_{\rm hard} \sim E/M^2$.}. The radiation amplitude is then given by the two diagrams of Fig.~\ref{fig:QEDlargexfrad}. 

%%%%%%%%%%%%%%%%%%%%%%%%%%%%%%%%%%%%%%%%%%%%%%%%%%%%%%%%%%%%%%%%%%%%%%%%%%%%%%
%%%%%%%%%%%%%%%%%%%%%%%%%%%%%%%%%%%%%%%%%%%%%%%%%%%%%%%%%%%%%%%%%%%%%%%%%%%%%

The radiation spectrum is thus given by the expression \eq{softspec} (or \eq{softspec1}), where $\vec{v}{\, '}$ is now the velocity of the outgoing muon. For small angle scattering, $\vec{v} \cdot \vec{v}{\, '} \simeq 1$. Denoting $\theta_s \ll 1$ the angle between $\vec{v}$ and $\vec{v}{\, '}$, and using $\theta_m^2 \equiv m^2/E^2 = 1-v^2 \ll 1$ and $\theta_M^2 \equiv M^2/E^2 \simeq 1-v'^2 \ll 1$, the relative velocity $\beta$ (see \eq{beta}) between the electron and muon can be approximated by
\be
\beta \simeq \sqrt{1-\frac{4\theta_m^2 \theta_M^2}{(\theta_M^2 + \theta_m^2 + \theta_s^2 )^2}} \, .
\label{beta-approx}
\ee
Since $M \gg m$, we have $\beta \to 1$. Using \eq{int-betatoone} we find 
\be
{\cal I}(v, v', \vec{v} \cdot \vec{v}{\, '}) \mathop{\simeq}_{\beta \to 1} \ln{\left( 1-\vec{v} \cdot \vec{v}{\, '} \right)} \simeq \ln{\left( \frac{\theta_M^2 + \theta_m^2 + \theta_s^2}{2} \right)} \, . 
\label{int-approx}
\ee
Thus, the interference of the two diagrams of Fig.~\ref{fig:QEDlargexfrad} is enhanced by a collinear logarithm, in contrast with the large angle scattering case discussed in section~\ref{sec:largept}. 

The radiation spectrum is obtained from \eq{softspec} by using \eq{col-log} and \eq{int-approx}. For $p$--$p$ collisions, 
\bea
\omega \left. \frac{dI}{d\omega} \right|_{\rm pp} &\simeq& \frac{\alpha}{\pi} \left[ \ln{\left(\frac{1}{\theta_m^2}\right)} + \ln{\left(\frac{1}{\theta_{M}^2}\right)} \right. \nn \\
&& \left. \ \ \ - 2 \ln{\left(\frac{1}{\theta_M^2+ \left. \theta_s^2 \right|_{\rm pp}} \right)}  \right] \, ,
\label{ppspec-largexf}
\eea
The first two terms stand for the DGLAP radiation of the electron and
muon respectively, and the third term is the interference. Note that
the interference term largely compensates the DGLAP terms. We easily
verify that the spectrum \eq{ppspec-largexf} arises from formation times $t_f \gg E/M_{\perp}^2$, \ie, $t_f \gg t_{\rm hard}$, as initially assumed. We stress that \eq{ppspec-largexf} is the spectrum associated to the hard process. Here, contrary to section~\ref{sec:largept}, it {\it cannot} be interpreted as the radiation spectrum of a well-identified charge, due to the presence of a large interference term.

In $p$--A collisions, the angle between the outgoing muon and incoming
electron tends to increase, due to transverse momentum broadening in
the target nucleus. For a nuclear target of sufficiently large size $L
\gg \lambda$ (with $\lambda$ the electron mean free path in the
target), transverse momentum broadening is given by the random walk
estimate $\Delta q_\perp^2 \sim (L/\lambda)\,\mu^2$, where $\mu$ is
the typical transverse exchange in a single scattering. Thus,
$\left. \theta_s^2 \right|_{\rm pA} = \left. \theta_s^2 \right|_{\rm
  pp} + \Delta \theta_s^2$, where  $\Delta \theta_s^2 = \Delta
q_\perp^2/E^2$ is the angular broadening. Formation times $t_f \gg
t_{\rm hard} \gg L$ being dominant (as in the $p$--$p$ scattering case), the spectrum is equivalent to that associated to a single {\it effective} scattering of transverse exchange $q_\perp^2 = \left. q_\perp^2 \right|_{\rm pp} +  \Delta q_\perp^2$. We focus on the limit $\Delta q_\perp^2 \ll \left. q_\perp^2 \right|_{\rm pp}$, so that the hard production process is not affected by soft rescatterings. The broadening of $\theta_s$ induces a slight modification of the interference term in $p$--A collisions, as compared to $p$--$p$ (see \eq{int-approx}), leading to the spectrum
\bea
\omega \left. \frac{dI}{d\omega} \right|_{\rm pA} &\simeq& \frac{\alpha}{\pi} \left[ \ln{\left(\frac{1}{\theta_m^2}\right)} + \ln{\left(\frac{1}{\theta_{M}^2}\right)} \right. \nn \\
&& \left. \ \ \ - 2 \ln{\left(\frac{1}{\theta_M^2+ \left. \theta_s^2 \right|_{\rm pA}} \right)}  \right] \, ,
\label{pAspec-largexf}
\eea
with again a large contribution from the interference term. 

The {\it medium-induced} spectrum is obtained by subtracting \eq{ppspec-largexf} from \eq{pAspec-largexf}, 
\bea
\omega \left. \frac{dI}{d\omega} \right|_{\rm ind} \simeq \frac{2 \alpha}{\pi}  \, \ln{\left(1+\frac{\Delta \theta_s^2}{\theta_M^2 + \left. \theta_s^2 \right|_{\rm pp}}\right)}  && \label{indspeclargexf3} \\
\simeq \frac{2 \alpha}{\pi}  \, \ln{\left(1+\frac{\Delta
    q_\perp^2}{M_\perp^2} \right)} \simeq \frac{2 \alpha}{\pi}  \,
\frac{\Delta q_\perp^2}{M_\perp^2}  \, . && \label{indspeclargexf3-b}
\eea

The expression \eq{indspeclargexf3-b} was obtained using \eq{softspec}, which arises from an exact angular integration. At high energy, all relevant angles ($\theta, \theta_s, \theta_m, \theta_M$) are small, and one could have as well worked in the small angle approximation from the beginning. In view of our generalization to the non-abelian situation in section~\ref{sec:quarkonium}, it is useful to mention how the expression \eq{indspeclargexf3} arises within this approximation. Expressing the square appearing in \eq{photonintensity} in the small angle limit, it is easy to check that the radiation spectrum associated to our small angle scattering process is given by \eq{ppspec}, with $\vec{J}_{\rm vac}$ replaced by the in-medium ``emission current'' 
\be
\vec{J}_{\rm med} = \frac{\vec{\theta}'_{\rm pA}}{\theta'^2_{\rm pA} + \theta_M^2} - \frac{\vec{\theta}}{\theta^2 + \theta_m^2}  \, .
\label{Jmedlargexf}
\ee
The second term arises from Fig.~\ref{fig:QEDlargexfrad}a (emission off the electron line) and the first, where $\vec{\theta}'_{\rm pA} = \vec{\theta} - \vec{\theta}_s$ with $\theta_s^2 =  \left. \theta_s^2 \right|_{\rm pA}$, from Fig.~\ref{fig:QEDlargexfrad}b (emission off the muon line). The medium-induced spectrum thus reads
\bea
\omega \left. \frac{dI}{d\omega} \right|_{\rm ind} \simeq \frac{\alpha}{\pi^2} \int d^2\vec{\theta} \, \left( \vec{J}_{\rm med}^{\,\, 2} - \vec{J}_{\rm vac}^{\,\, 2} \right) \hskip 2cm && 
\label{indspeclargexf3-smallangle-a} \\ 
= - \frac{2\alpha}{\pi^2} \int \frac{d^2\vec{\theta}}{\theta^2 + \theta_M^2}  \left[ \frac{\vec{\theta} \cdot \vec{\theta}'_{\rm pA}}{\theta'^2_{\rm pA}}  - ({\rm vac})  \right] \, , \ \ \ \ \ \  &&
\label{indspeclargexf3-smallangle}
\eea
where we set $m \to 0$ and the vacuum contribution is obtained by
replacing $\left. \theta_s^2 \right|_{\rm pA} \to \left. \theta_s^2 \right|_{\rm pp}$. From \eq{azimint} we have 
\be
\label{azimint2}
\int \frac{d\phi}{2\pi} \, \frac{\vec{\theta} \cdot \vec{\theta}'}{\theta'^2} = \Theta(\theta^2 - \theta^2_s)  \, ,
\ee
and we arrive at 
\be
\omega \left. \frac{dI}{d\omega} \right|_{\rm ind} \simeq
\frac{2\alpha}{\pi} \int_{\left. \theta_s^2 \right|_{\rm
    pp}}^{\left. \theta_s^2 \right|_{\rm pA}}
\frac{d\theta^2}{\theta^2 + \theta_M^2}  \, ,
\label{qed-ind-spectrum}
\ee
from which the expression \eq{indspeclargexf3} follows.

The medium-induced spectrum \eq{qed-ind-spectrum} is dominated by
formation times 
\be
t_f \sim \frac{1}{\omega(\theta^2 + \theta_M^2)} \simeq
\frac{1}{\omega(\left. \theta_s^2 \right|_{\rm pp} + \theta_M^2)}
  = \frac{E^2}{\omega M_{\perp}^2} \gg \frac{E}{M_{\perp}^2} \, .
\ee
Hence, $t_f \gg t_{\rm hard} \gg L$, justifying our initial assumption. Contrary to the case studied in section~\ref{sec:largept}, large photon formation times do not cancel out and dominate in the medium-induced spectrum. This justifies approximating the radiation spectrum as that associated to a single effective scattering, and using \eq{softspec} throughout the discussion. The induced spectrum arises solely from the (induced) interference term, and cannot be identified with the induced spectrum of a given charged particle, either the electron or the muon. To stress this point, let us note that if the incoming (outgoing) energetic particle carries a charge $e$ ($e'$) in units of the electron charge, the medium-induced spectrum will be given by \eq{indspeclargexf3} multiplied by $e e'$. 

The induced spectrum \eq{indspeclargexf3} is very similar to the Bethe-Heitler (BH) radiation spectrum off an asymptotic muon crossing a nucleus. The latter is obtained from \eq{softspec1} by setting $m=M$ in \eq{beta-approx}. For $\theta_s =0$, we have $\beta =0$ and the BH spectrum vanishes. For $\theta_s \neq 0$,  
\be
\beta = \sqrt{1-\frac{1}{(1+a/2)^2}} \ \ \ ; \ \ \ a \equiv \frac{\theta_s^2}{\theta_M^2} \, ,
\ee
and the function ${\cal R}(\beta)$ appearing in \eq{softspec1} and defined by \eq{Rofv} is very well approximated by $\ln{(1+a/3)}$, which has the same limiting behaviours as ${\cal R}(\beta)$ when $a \ll 1$ and $a \gg 1$. The soft photon radiation spectrum off an asymptotic massive charge thus reads~\cite{Peigne:2008wu}
\be
\omega \left. \frac{dI}{d\omega} \right|_{\rm muon} \simeq \frac{2 \alpha}{\pi} \ln{\left(1+\frac{\theta_s^2}{3 \theta_M^2} \right)} \simeq \frac{2 \alpha}{3 \pi}  \, \frac{\Delta q_\perp^2}{M^2}  \, .
\label{specasymmuon2}
\ee
In the small angle approximation, the result \eq{specasymmuon2} would be obtained from \eq{Jmedlargexf} and \eq{indspeclargexf3-smallangle-a} by setting $M=m$,
\bea
\omega \left. \frac{dI}{d\omega} \right|_{\rm muon} \simeq \frac{\alpha}{\pi^2} \int d^2\vec{\theta} \, \left( \frac{\vec{\theta}{\, '}}{\theta'^2 + \theta_M^2} - \frac{\vec{\theta}}{\theta^2 + \theta_M^2} \right)^2 && \nn \\ 
= \frac{2 \alpha}{\pi^2} \int d^2\vec{\theta} \, \frac{\vec{\theta}}{\theta^2 + \theta_M^2} \left( \frac{\vec{\theta}}{\theta^2 + \theta_M^2} - \frac{\vec{\theta}'}{\theta'^2 + \theta_M^2} \right) \, . \ \ \ \ \ && 
\label{specasymmuon1-smallangle}
\eea
Performing the angular integral one recovers \eq{specasymmuon2}. 

The BH spectrum \eq{specasymmuon2} is qualitatively similar to the radiation spectrum \eq{indspeclargexf3} associated to the hard production process of Fig.~\ref{fig:QEDlargexf}. In particular it is dominated by large formation times $t_f \gg L$. It is worth noting that the radiation spectrum of an asymptotic charge can be obtained from the induced spectrum \eq{inducedspec1}, \eq{inducedspec2} of a charge produced in a medium by moving the production point to the far past, $L_0 \to \infty$. This is another way to understand why large formation times contribute in the present case. The typical formation time contributing to the induced loss of a suddenly accelerated particle (see section~\ref{sec:largept}) is $t_f \sim L_0 \sim L$. When $L_0$ increases at fixed $L$, $t_f$ increases to larger and larger values, and eventually saturates at $t_f \sim 1/(\omega \theta_M^2)$. 

Integrating the energy spectrum \eq{indspeclargexf3-b} up to $\omega \sim E$, we obtain the medium-induced {\it radiated energy} (rather than {\it energy loss}) associated to the small angle scattering process,
\be
\Delta E_{\rm ind} \sim \alpha \cdot \frac{\Delta q_\perp^2}{M_\perp^2} \cdot E \sim \alpha \cdot \frac{L \mu^2}{\lambda M_\perp^2} \cdot E \, ,
\label{largexfelectronloss}
\ee 
to be contrasted with the result \eq{t=0electronloss} for large angle scattering.
We note that the parametric dependence of the {\it radiated energy} \eq{largexfelectronloss} is formally the same as that of the energy loss of an asymptotic charge of mass $M_\perp$, see \eq{specasymmuon2}. 

To summarize this section: {\it In the case of small-angle scattering, the medium-induced photon radiation spectrum associated to the hard process arises from the interference between initial and final state radiation. As such, it cannot be identified with the medium-induced spectrum of well-defined charged particles. Nevertheless, the spectrum is qualitatively the same as that of an asymptotic massive charge crossing the medium. In particular, it arises from formation times $t_f \gg L$, resulting in a medium-induced radiated energy scaling as $E$.}

\section{Large $\xf$ quarkonium production}
\label{sec:quarkonium}

In this section we consider the QCD process of quarkonium (hadro)production at large-$\xf$, in $p$--$p$ and $p$--A collisions. We discuss the medium-induced radiation spectrum associated to the hard process, given by the difference between the soft gluon radiation spectra associated to vacuum ($p$--$p$) and in-medium ($p$--A) production. We will show that the features of gluon radiation are similar to those obtained in the QED model of section~\ref{sec:largexf} (despite some difference in the parametric behaviour of the photon and gluon energy spectra). In particular, the medium-induced spectrum arises from large gluon formation times, leading to a medium-induced radiated energy scaling as the quarkonium energy. 

\subsection{Model for large-$\xf$ quarkonium hadroproduction}
\label{sec:IIIA}

In order to single out the main features of large-$\xf$ quarkonium
production, we use several simplifying assumptions. As mentioned in
the Introduction, we focus on the partonic process $gg \to \QQ$, where
the final quark and antiquark momenta are similar and
quasi-collinear. At large $x_F = x_1 - x_2 \simeq x_1$ and in the
target rest frame, the incoming gluon splits into the $\QQ$ pair, to
which it transfers most of its energy $E$, equally shared between the
quark and antiquark. The coherence time of the hard process is given 
by \eq{lhard} with $M_X=M_{\perp}$, $M_{\perp} \gg \Lambda_{\rm QCD}$ denoting the quarkonium transverse mass. Thus, $t_{\rm hard} \sim E/M_{\perp}^2$, corresponding to the $g \to \QQ$ fluctuation time in the target rest frame.  

In the following we will not have to specify the precise mechanism for the transition between the $\QQ$ pair and the quarkonium bound state. We will however assume that the $\QQ$ pair remains a color octet for a time $t_{\rm octet} \gg E/M_{\perp}^2$. This is a reasonable assumption for the Color Octet Mechanism (COM)~\cite{Bodwin:1994jh}, or in the Color Evaporation Model (CEM)~\cite{Fritzsch:1977ay} for quarkonium production. In the Color Singlet Model (CSM)~\cite{Chang:1979nn}, the quantum numbers of the quarkonium bound state (in particular its color neutrality) are fixed in the perturbative process, \ie, $t_{\rm octet}^{\rm CSM} \sim t_{\rm hard} \sim E/M_{\perp}^2$ instead of $t_{\rm octet} \gg E/M_{\perp}^2$. Our results might nevertheless also have implications on quarkonium production within the CSM, see section~\ref{sec:heavyflav} for a discussion of this point.  

In section~\ref{sec:IIIB} below, we will study the medium-induced
gluon radiation spectrum associated to large-$\xf$ quarkonium hadroproduction in the $\omega$ domain where
\be
t_{\rm hard} \sim \frac{E}{M_{\perp}^2} \ll t_f \sim \frac{1}{\omega
  (\theta^2 + \theta_M^2)} \ll t_{\rm octet} \, ,
\label{tfassumption}
\ee
the first inequality implying that radiation does not probe the hard $gg \to \QQ$ partonic process, and the se\-cond that the $\QQ$ pair remains color octet during the overall radiation process. We will also assume that the hard scale $M_\perp = \sqrt{M^2 +q_\perp^2}$ is the largest scale after the quarkonium energy $E$, where $q_\perp$ is the transverse momentum of the $\QQ$ pair. In particular, the radiated gluon has $k_\perp \ll M_\perp, q_\perp$. Under those conditions, the radiation does not affect the kinematics of the hard process, and does not probe the transverse size $\sim \morder{1/M_{\perp}}$ of the $\QQ$ pair \footnote{Strictly speaking, the size of the $\QQ$ pair is $\sim \morder{1/M}$ only during the time $t_{\rm hard} \sim E/M^2$, after which it increases up to the quarkonium radius $\sim \morder{1/(\alpha_s(M) M)}$. Thus, in order to treat the $\QQ$ pair as pointlike during the time interval $t_{\rm hard} \ll t_f \ll t_{\rm octet}$ (at least in the CEM and COM), a sufficient condition is $k_\perp \ll \alpha_s(M) M$ instead of $k_\perp \ll M$. However, the distinction between the scales $M$ and $\alpha_s(M) M$ is irrelevant in our discussion.}, and the latter thus appears as a {\it compact} color octet state, \ie, as a (massive) gluon~\cite{Hoyer:1997yf}. 

\begin{figure}[h]
\centering
\includegraphics[scale=1.4]{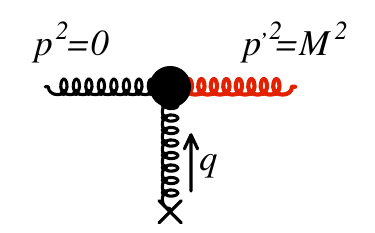}
\caption{QCD model for quarkonium hadroproduction at large $\xf$. The ``massive gluon'' turns into a color-neutral object (and eventually, into a quarkonium) on longer time scales in the CEM and COM.}
\label{fig:effQQamp}
\end{figure}

For our purpose, the quarkonium production process is effectively equivalent to the process depicted in Fig.~\ref{fig:effQQamp}, where an energetic ``massive gluon'' is produced in the hard scattering of an incoming (massless) gluon. We stress that the incoming gluon should not be coherent with the other partons of the projectile proton for our perturbative treatment to be meaningful. This requires the typical transverse momenta to probe the proton size, $k_\perp, q_\perp \gg \Lambda_{\rm QCD}$. 

%%%%%%%%%%%%%%%%%%%%%%%%%%%%%%%%%%%%%%%%%%%%%%%%%%%%%%%%%%%%%%%%%%%%%%%%%%%%%%
%%%%%%%%%%%%%%%%%%%%%%%%%%%%%%%%%%%%%%%%%%%%%%%%%%%%%%%%%%%%%%%%%%%%%%%%%%%%%

\subsection{Medium-induced radiation spectrum}
\label{sec:IIIB}

We now want to derive the medium-induced radiation spectrum associated to large-$\xf$ quarkonium production, focusing on the case of a medium of size $L \gg \lambda$, where $\lambda$ is the gluon mean free path between successive elastic scatterings in the target nucleus. As in the QED case studied in section~\ref{sec:largexf}, we anticipate that the spectrum arises from large (gluon) formation times, see \eq{tfassumption}, in particular $t_f \gg L$ at large $E$. Under those conditions, gluon radiation does not probe the hard process, and only emission vertices before or after the hard blob of Fig.~\ref{fig:effQQamp} have to be considered. Similarly to Ref.~\cite{Peigne:2008wu}, we use semi-heuristic arguments to derive the radiation spectrum. 

The emission vertex off the final gluon line, obtained by factoring out the amplitude {\it without} radiation (see Fig.~\ref{fig:effQQamp}) from the amplitude {\it with} radiation, has the same structure as in QED (up to some implicit color factor), 
\be
\vec{J}_{\rm final} \sim - \frac{\vec{\theta}{\, '}}{\theta'^2 + \theta_M^2} \ \ \ ; \ \ \ \vec{\theta}{\, '} = \vec{\theta} - \vec{\theta}_s = \vec{\theta} - \frac{\vec{q}_\perp}{E} \, .
\label{final-rad}
\ee
The term $\theta_M^2$ in the denominator arises from the fact that the pointlike color octet $\QQ$ pair (\ie, the final ``gluon'' in Fig.~\ref{fig:effQQamp}) has a mass $M$, and $\vec{\theta}{\, '}$ is the emission angle with respect to the final ``gluon'' direction ($\vec{\theta}_s = \vec{q}_\perp /E$ is the energetic gluon scattering angle).  

\begin{figure}[h]
\centering
\includegraphics[scale=1.3]{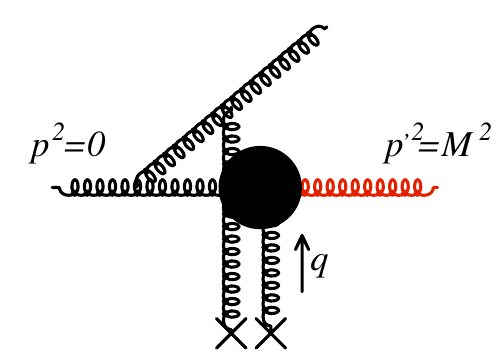}
\caption{Generic amplitude for gluon radiation in the  model of Fig.~\ref{fig:effQQamp}. In QCD the radiated gluon also rescatters in the medium.}
\label{fig:genericamp}
\end{figure}

The emission vertex off the initial (massless) gluon line is a priori more difficult to extract, since the radiated gluon can rescatter in the target. Let us consider the process in a $p$--A collision. A generic diagram for the radiation amplitude is shown in Fig.~\ref{fig:genericamp}. Compared to the process in $p$--$p$ scattering, the radiated gluon suffers some transverse momentum broadening $\Delta q_\perp^2 \sim (L/\lambda) \, \mu^2$ \footnote{As noted in section~\ref{sec:IIIA}, in order to treat the $\QQ$ pair as a pointlike color octet, we need all scales but $E$ to be smaller than $M$, in particular $(L/\lambda) \, \mu^2 \ll M^2$ in what follows.}. The emission vertex can be obtained heuristically by noting that just after the emission, the angle between the radiated and initial gluon momenta is $\vec{\theta}{\, ''} = \vec{\theta} - \vec{\theta}_g$, where $\vec{\theta}_g = \Delta \vec{q}_\perp /\omega$ is the {\it rescattering} angle of the radiated gluon in the target~\cite{Peigne:2008wu}. The fact that in Fig.~\ref{fig:genericamp} the radiated gluon might couple to the rescattering ($\Delta q_\perp$) gluon but not to the hard ($q_\perp$) exchanged gluon arises from our assumption of {\it soft} gluon radiation, which should not affect the kinematics of the hard production process, and thus factorize from it. Clearly, our picture requires $k_{\perp}, \Delta q_{\perp} \ll q_{\perp}$ to be valid. Thus, effectively,
\be
\vec{J}_{\rm initial} \sim \frac{\vec{\theta}{\, ''}}{\theta''^2 } \ \ \ ; \ \ \ \vec{\theta}{\, ''} = \vec{\theta} - \vec{\theta}_g  = \vec{\theta} - \frac{\Delta \vec{q}_\perp}{\omega}\, . 
\label{initial-rad}
\ee

%%%%%%%%%%%%%%%%%%%%%%%%%%%%%%%%%%%%%%%%%%%%%%%%%%%%%%%%%%%%%%%%%%%%%%%%%%%%%%
%%%%%%%%%%%%%%%%%%%%%%%%%%%%%%%%%%%%%%%%%%%%%%%%%%%%%%%%%%%%%%%%%%%%%%%%%%%%%

The medium-induced spectrum associated to large-$\xf$ quarkonium production is obtained by multiplying \eq{final-rad} and \eq{initial-rad} \footnote{The contributions from DGLAP radiation cancel in the {\it medium-induced} radiation spectrum, as in the QED model of section~\ref{sec:largexf}.}, integrating over $d^2\vec{\theta}$, and then subtracting the ``vacuum'' contribution (formally obtained by $\Delta \vec{q}_\perp \to \vec{0}_\perp$) corresponding to $p$--$p$ scattering. After a simple shift in the angular integration variable $\vec{\theta}$ we obtain:
\bea
\omega \left. \frac{dI}{d\omega} \right|_{\rm ind} &\sim& - \frac{N_c \alpha_s}{\pi^2} \int \frac{d^2\vec{\theta}}{\theta^2 + \theta_M^2} \left[ \frac{ \vec{\theta} \cdot (\vec{\theta} + \vec{\theta}_s - \vec{\theta}_g)}{(\vec{\theta} + \vec{\theta}_s - \vec{\theta}_g)^2} - ({\rm vac}) \right]  \nn \\
\label{indspec-quarkonium}
\eea
The latter expression is similar to the QED expression \eq{indspeclargexf3-smallangle}, up to the replacement $\vec{\theta}_s \to \vec{\theta}_s - \vec{\theta}_g$. Using \eq{azimint2}, the angular integration gives:
\bea
\omega \left. \frac{dI}{d\omega} \right|_{\rm ind} 
&\sim& \frac{N_c \alpha_s}{\pi} \int_{\left. (\vec{\theta}_s - \vec{\theta}_g)^2 \right|_{\rm pp}}^{\left. (\vec{\theta}_s - \vec{\theta}_g)^2 \right|_{\rm pA}} \frac{d \theta^2}{\theta^2 + \theta_M^2} \label{indspec-theta-int} \\  
&\sim& \frac{N_c \alpha_s}{\pi}  \, \ln{\left(1+\frac{\Delta q_\perp^2 \, E^2}{M_\perp^2 \, \omega^2}\right)} \, . 
\label{indspec-quarkonium-final}
\eea 
In order to obtain the second line, we approximated
\bea
\left. (\vec{\theta}_s - \vec{\theta}_g)^2 \right|_{\rm pA} &\simeq& \left. \theta_s^2 \right|_{\rm pA} + \left. \theta_g^2 \right|_{\rm pA} \simeq \frac{q_\perp^2}{E^2} + \frac{\Delta q_\perp^2}{\omega^2} \hskip 5mm \label{appr1} \\ 
\left. (\vec{\theta}_s - \vec{\theta}_g)^2 \right|_{\rm pp} &\simeq& \left. \theta_s^2 \right|_{\rm pp} \simeq \frac{q_\perp^2}{E^2} \label{appr2} 
\eea

Comparing to the QED result \eq{indspeclargexf3-b}, we observe that the QCD spectrum \eq{indspec-quarkonium-final} involves a new scale,
\bea
\omega \left. \frac{dI}{d\omega} \right|_{\rm ind} &\sim& \frac{N_c \alpha_s}{\pi}  \, \ln{\left(1+\frac{\hat{\omega}^2}{\omega^2}\right)} \label{spectrum-omegahat} \\
\hat{\omega} &\equiv&  \frac{\sqrt{\Delta q_\perp^2}}{M_\perp} \, E \ll E \, , 
\label{omegahat}
\eea
above which the spectrum is suppressed as $\sim 1/\omega^2$. 

The medium-induced {\it radiated energy} is obtained by integrating the spectrum \eq{spectrum-omegahat} over $\omega$, up to $\omega \sim E$. However, due to fast convergence for $\omega > \hat{\omega}$, the integral is well approximated by replacing $E$ by infinity. We find:
\be
\left. \Delta E \right|_{{\rm ind,\ large}\ x_F} \sim N_c \alpha_s \, \hat{\omega} \sim  N_c \alpha_s \frac{\sqrt{\Delta q_\perp^2}}{M_\perp} \cdot E \, .
\label{quarkoniumloss}
\ee
The medium-induced radiated energy arises from gluon energies $\omega
\sim \hat{\omega} \ll E$, and thus from gluon formation times $t_f
\sim E^2/(\omega M_{\perp}^2) \gg E/M_{\perp}^2$, as initially
assumed. Note also that $k_\perp \simeq \omega \theta \sim
\hat{\omega} M_{\perp} /E \ll M_{\perp}$ and thus the ra\-dia\-ted gluon cannot probe the transverse size $\sim \morder{1/M_{\perp}}$ of the heavy $\QQ$ pair.

The medium-induced radiated energy \eq{quarkoniumloss} scales as the quarkonium energy, due to a dominant contribution from large gluon formation times. Similarly to the QED case, it cannot be attributed to any well-identified parton, since it arises from the interference between emission vertices off an incoming massless gluon and an outgoing massive pointlike color octet. However, \eq{quarkoniumloss} exhibits the same parametric dependence as the radiative loss of an asymptotic color charge of ``mass'' $M_{\perp}$ undergoing a single effective scattering of momentum transfer $\Delta q_\perp^2$~\cite{Peigne:2008wu}. In particular, it behaves as $1/M_{\perp}$ rather than $1/M_{\perp}^2$ as in QED (see \eq{largexfelectronloss}). 

Finally, in view of phenomenological applications, let us determine the validity range of the spectrum \eq{spectrum-omegahat}. First, the spectrum was derived in the soft gluon approximation, $\omega \ll E$. Second, and most importantly, in QCD our perturbative derivation is meaningful provided $k_\perp > \Lambda_{\rm QCD}$, as already mentioned in section \ref{sec:IIIA}. Using $k_\perp \simeq \omega \theta$ and the approximations \eq{appr1}, \eq{appr2}, the spectrum \eq{indspec-theta-int} is reexpressed as 
\be
\omega \left. \frac{dI}{d\omega} \right|_{\rm ind} \sim \frac{N_c \alpha_s}{\pi} \int_{{\rm Max}(x^2 q_\perp^2, \Lambda_{\rm QCD}^2)}^{x^2 q_\perp^2 + \Delta q_\perp^2} \frac{d k_\perp^2}{k_\perp^2 + x^2 M^2} \, ,
\label{indspec-kperp-int} 
\ee
where the constraint $k_\perp>\Lambda_{\rm QCD}$ is now taken into
account, and $x = \omega/E$. Note that $\Delta q_\perp^2\sim (L/\lambda)\,\mu^2\gg\Lambda_{\rm QCD}^2$ for large nuclei. Approximating ${\rm Max}(x^2 q_\perp^2, \Lambda_{\rm QCD}^2) \sim x^2 q_\perp^2 +\Lambda_{\rm QCD}^2$, we find
\bea
\omega \left. \frac{dI}{d\omega} \right|_{\rm ind} &\sim& \frac{N_c \alpha_s}{\pi}\, \ln{\left(\frac{1 + \hat{\omega}^2/\omega^2}{1 + \omega_0^2/\omega^2} \right)} \label{corrected-spectrum} \\
\omega_0 &\equiv& \frac{\Lambda_{\rm QCD}}{M_\perp} \, E < \hat{\omega} \, .
\eea 

%%%%%%%%%%%%%%%%%%%%%%%%%%%%%%%%%%%%%%%%%%%%%%%%%%%%%%%%%%%%%%%%%%%%%%%%%%%%%%%%%%%%%%%%%%%%%%%%%%%
\begin{figure}[t]
\centering
\includegraphics[scale=.27]{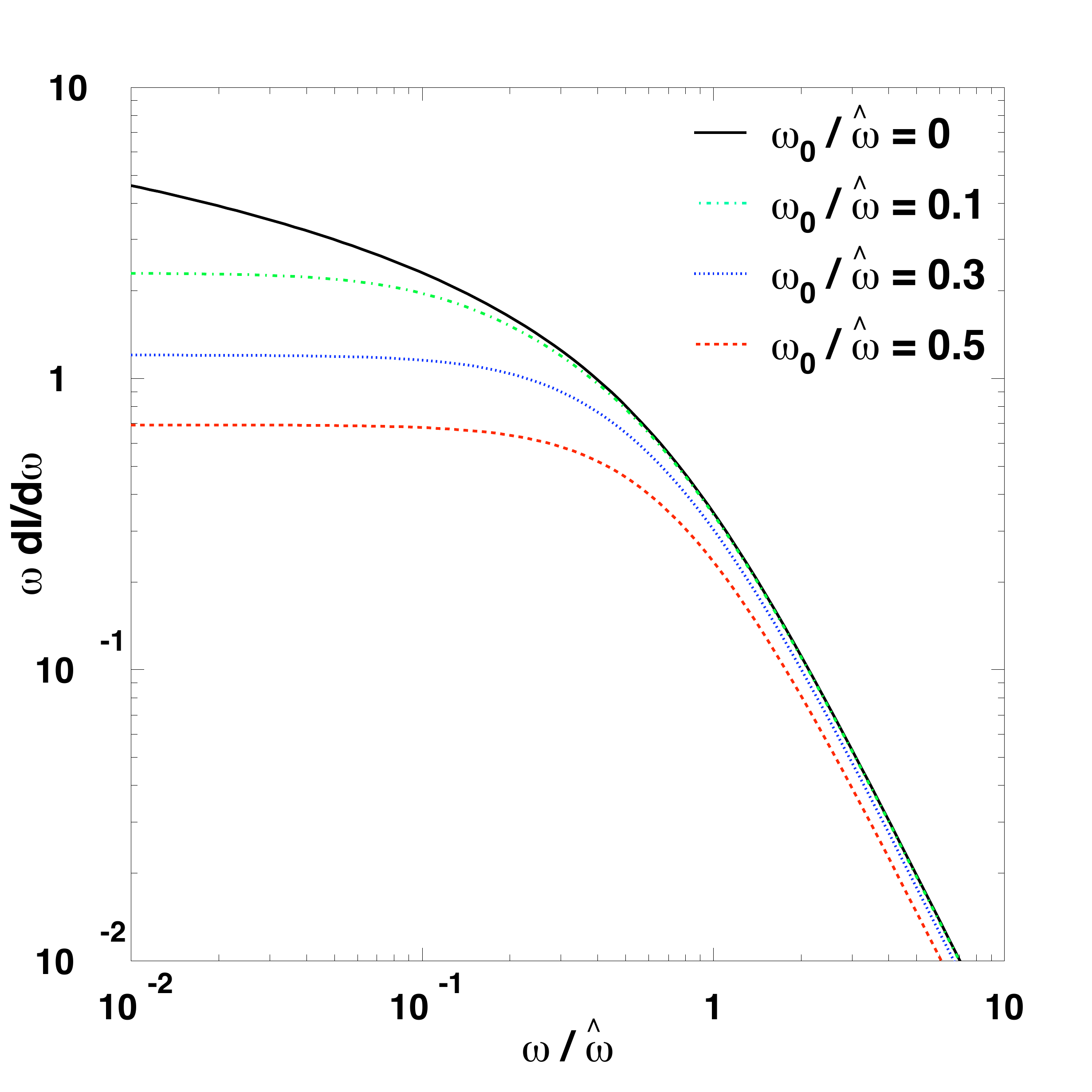}
\caption{Medium-induced soft gluon radiation spectrum \eq{corrected-spectrum} associated to large $\xf$ quarkonium production, for various values
of $\omega_0/\hat{\omega}$.
We chose $N_c \alpha_s/\pi=1/2$ to draw the figure.}
\label{fig:schematic}
\end{figure}
%%%%%%%%%%%%%%%%%%%%%%%%%%%%%%%%%%%%%%%%%%%%%%%%%%%%%%%%%%%%%%%%%%%%%%%%%%%%%%%%%%%%%%%%%%%%%%%%%%%
The spectrum is shown in Fig.~\ref{fig:schematic} from which we see that Eq.~\eq{spectrum-omegahat} is valid down to $\omega \sim \omega_0$,
below which it becomes smoother and the logarithm saturates at $\ln{(\hat{\omega}^2/\omega_0^2)} = \ln{(\Delta
  q_\perp^2/\Lambda_{\rm QCD}^2)} \simeq \ln{(L/\lambda)}$.

With the constraint $k_\perp > \Lambda_{\rm QCD}$, the average medium-induced radiated energy now reads
\be
\left. \Delta E \right|_{{\rm ind,\ large}\ x_F} \sim  N_c \alpha_s \,
\left( \hat{\omega}-\omega_0 \right) \, .
\ee
The result \eq{quarkoniumloss} thus receives a reduction factor
\be
1-\frac{\omega_0}{\hat{\omega}}  \simeq 1-\sqrt{\frac{\lambda}{L}
  \cdot \frac{\Lambda_{\rm QCD}^2}{\mu^2}} \, , 
\label{corrected-quarkoniumloss}
\ee
which is about $0.5$ when $L / \lambda = 4$ and $\mu \simeq \Lambda_{\rm QCD}$.

\section{Comparing different processes}
\label{sec:disc}

The main message of our study can be stated as follows. In a hard process involving incoming and outgoing energetic charges (which do not have to be identical) being quasi-collinear in the rest frame of the medium, the associated, {\it medium-induced} radiated energy --~obtained by subtracting the contribution in $p$--$p$ from that in $p$--A scattering~-- arises from the interference between the emission amplitudes off the initial and final charges, and is dominated by large gluon formation times. The medium-induced radiated energy is similar to the medium-induced {\it energy loss} of an asymptotic charge, in particular it scales as the energy, which might have important consequences on the phenomenology.

In this section we discuss various hard processes off nuclear targets and examine for each of them whether the medium-induced energy loss is similar to that of an asymptotic charge or not. The quantitative effects of medium-induced gluon radiation on large-$\xf$ quarkonium, open heavy flavour, and light hadron hadroproduction will be addressed in a future study. Since the gluon radiation spectra associated to these processes are similar (they only differ by the precise value of the scales $\hat{\omega}$
and $\hat{\omega}_0$ appearing in \eq{corrected-spectrum}), we anticipate that similar (and large) nuclear suppressions will be obtained in those cases.

\subsection{DIS and photoproduction}
\label{sec:dis}

In DIS on nuclei, the medium-induced energy loss is that of the parton knocked by the virtual photon. In this case, the expected energy loss is that of a charged particle produced at $t=0$, as assumed in various phenomenological analyses~\cite{Arleo:2003jz}.

The situation is a little more complex in photoproduction reactions. As long as the initial photon directly couples to the quark in the target nucleus (at leading order), the process resembles DIS and the energy loss is that of the struck quark. In photoproduction however, the hadronic structure of the photon can also be resolved: in this case a colored parton stemming from the photon participates to the hard scattering dynamics, making resolved photoproduction very similar to hadroproduction. Since the time scale to resolve the parton inside the photon is typically much larger than the time at which the hard process occurs, one could expect an interference between the emission amplitudes off the initial and final-state parton, leading to a medium-induced energy loss (or, more accurately, radiated energy) scaling like the parton energy. In this respect, it would be extremely valuable to investigate the nuclear dependence of forward jet and hadron production in direct {\it vs.} resolved photoproduction in nuclei.

\subsection{Drell-Yan production}
\label{sec:dy}

As already mentioned, in the absence of color charge in the partonic final state, the energy loss in Drell-Yan production is expected to be that of a suddenly decelerated parton, \ie, independent of its energy \footnote{This statement is at variance with Ref.~\cite{Vitev:2007ve}, where a medium-induced energy loss scaling as $E$ is found in DY production, in apparent contradiction with the Brodsky-Hoyer bound (see \eq{constraint}) which should apply in this case.}, as is the case in large angle scattering (section~\ref{sec:largept}). The consequence is that the effects of energy loss should play almost no role in Drell-Yan production in high-energy $p$--A collisions, unlike what is assumed in the model by Gavin and Milana~\cite{Gavin:1991qk}.

The energy loss scaling as $E$ would come into play only if {\it another} energetic charged particle is produced in the final-state, in association with the virtual photon. Such a situation occurs in DY+jet production in $p$--A or A--A collisions, with the jet produced at large rapidity.

\subsection{Heavy-flavour hadroproduction}
\label{sec:heavyflav}

In the case of large-$\xf$ quarkonium hadroproduction, our result \eq{quarkoniumloss} holds provided gluon radiation has time to be formed before the $\QQ$ pair turns color singlet, see \eq{tfassumption}. As we mentioned in section~\ref{sec:IIIA}, this assumption is justified in the CEM and COM (provided color octet contributions to quarkonium production dominate in the latter case). As a consequence, one could expect in this case a different nuclear dependence of quarkonium hadroproduction at large $\xf$ and quarkonium photoproduction at large $z$ (see section~\ref{sec:dis}). 

However, we should recall that the precise dynamics of quarkonium production is still unknown. Indeed, no proposed model can explain all features of the data on quarkonium production. Although the CSM alone seems to be ruled out by hadroproduction data, it is worth keeping in mind that color singlet contributions might not be negligible in some kinematical regions~\cite{Lansberg:2008gk}. 

We have to stress that our main result \eq{spectrum-omegahat} (or more accurately \eq{corrected-spectrum}) does not apply to such color singlet contributions, for which $t_{\rm octet} \sim t_{\rm hard} \sim E/M_{\perp}^2$, in contrast to our assumption \eq{tfassumption}. In the CSM, the outgoing color octet $\QQ$ pair is too short-lived to allow for our interference contribution to gluon radiation. However, depending on the quarkonium quantum numbers, the CSM may require the quarkonium bound state to be produced {\it in conjunction} with a hard gluon, as for instance in $J/\psi$ production. In this case, the hard process looks like small angle scattering of an energetic color charge -- even though the final charge (the hard gluon) is distinct from the triggered $\QQ$ pair. The medium-induced radiation spectrum \eq{corrected-spectrum}, with $E$ interpreted as the energy of the hard gluon (and $M$ set to zero in the expression of $M_{\perp}$), might thus indirectly affect the $J/\psi$ production rate. In other cases, like $\chi_{c2}$ production in the CSM, where no associated hard gluon is required (at leading order in $\alpha_s$), the radiation spectrum \eq{corrected-spectrum} will not apply. 

In the COM the $\chi_{c2}$ state --~which radiative decays contribute quite significantly to $\jpsi$ production~-- is produced predominantly as a color singlet object. Therefore, we do not expect our result \eq{quarkoniumloss} to apply to this state, leading to a much smaller $\chi_{c2}$ suppression (as compared to $\jpsi$) at large $\xf$. This prediction, which could be tested at fixed-target experiments at Fermilab~\cite{Vogt:2001ky}, is at variance with other nuclear effects such as nuclear PDF effects or intrinsic charm which do not depend on the identity of the final charmonium state~\cite{Vogt:1999dw}. Another prediction can be made regarding the $\Upsilon$ suppression. From the mean radiated energy \eq{quarkoniumloss}, $\Delta E\sim 1/M_{\perp}$, the suppression of $\Upsilon$ production, for which no measurements have been performed yet at large $\xf$, is expected to be less pronounced than that of $\jpsi$.

We expect the medium-induced radiation spectrum \eq{corrected-spectrum} to be valid in {\it open} heavy flavour hadroproduction at large $\xf$. In particular, the fact that in this process the final energetic heavy quark does not have the same color charge as the octet $\QQ$ pair in quarkonium production does not affect our conclusions. As mentioned in the Introduction, the similar suppression of $\jpsi$ production and single muon (coming from $D$-decays) production in $p$--A collisions as a function of rapidity~\cite{e866opencharm} supports this picture.

\subsection{Light-hadron hadroproduction}

Unlike DIS, Drell-Yan and heavy-flavour production, in the case of {\it light} hadron production at large $\xf$ there is no natural hard scale, except if the final hadron transverse momentum in a $p$--$p$ collision is large enough, $p_{\perp}^{\rm had} \sim \left. q_\perp\right|_{\rm pp} \gg \Lambda_{\rm QCD}$, in which case a perturbative description might hold. Regarding the expected medium-induced radiation spectrum, different cases might occur depending on the actual definition of the observable:
\begin{itemize}
\item[(i)] the suppression is defined as the (normalized) ratio of the light hadron rate in $p$--A (or A--A) collisions over $p$--$p$ collisions at the same, {\it fixed} ${\left. q_\perp^2\right|_{\rm pA}}={\left. q_\perp^2\right|_{\rm pp}}$. In this case, the interference term in the induced radiation spectrum vanishes from the actual definition of the observable, see the QED expressions Eqs.~(\ref{intro-ppspec-largexf})  and~(\ref{intro-pAspec-largexf}), or set $\Delta q_\perp^2 = 0$ in the gluon radiation spectrum \eq{indspec-quarkonium-final}. There will nevertheless be some energy loss effect due to rescattering in the medium. However, in this case the energy loss will be dominated by the radiation of gluons which resolve the medium (\ie, with a formation time of the order of the medium size $L$) and will not scale as the parton energy; 
\item[(ii)] the light-hadron suppression in $p$--A or A--A collisions is defined as the (normalized) ratio of the light-hadron production rate {\it integrated} over all transverse momenta (with an appropriate infrared cut-off). In that case, ${\left. q_\perp^2\right|_{\rm pA}}={\left. q_\perp^2\right|_{\rm pp}}$ is no longer required, leading to a potentially large energy loss proportional to the parton energy. The induced gluon spectrum is then obtained by setting $M=0$, \ie, $M_\perp \to q_\perp$ in \eq{corrected-spectrum}.
\end{itemize}

In the case of the BRAHMS experiment at RHIC~\cite{Arsene:2003yk}, the quenching factor for all charged hadrons in $d$--Au compared to $p$--$p$ collisions at $\sqrt{s}=200$~GeV/nucleon and at forward rapidity ($0\leq y \lesssim 4$) is measured at fixed hadron transverse momentum $\pt$, which therefore looks like case (i) above. However note that the final parton transverse momentum is smeared because of the fragmentation process in QCD. Whenever the {\it typical dispersion} of the parton transverse momentum is large (as compared to the momentum broadening $\Delta q_\perp^2$), we no longer have  ${\left. q_\perp^2\right|_{\rm pA}}={\left. q_\perp^2\right|_{\rm pp}}$ and the energy loss might scale with the parton energy $E\propto \exp y$. Another reason why comparing at fixed $p_{\perp}^{\rm had}$ might nevertheless induce an energy loss similar to that of asymptotic partons is the binning used in the experiment. As long as the typical size of the $\pt$-bin is large as compared to the nuclear broadening, this case is similar to an {\it integrated} distribution over the transverse momenta, \ie, to case (ii).

The NA49 experiment at SPS also measured the $\xf$-differential production cross section of light hadrons in $p$--$p$ and $p$--A collisions and integrated over all transverse momenta~\cite{Fischer:2002qp}, which corresponds to the case (ii) above. The strong suppression observed at large $\xf$ therefore appears qualitatively consistent with our expectation~\footnote{Note however that no minimal $\pt$-cut has been applied, making these data possibly sensitive to non-perturbative dynamics.}.

\section{Outlook}
\label{sec:outlook}

As already mentioned, the notion of {\it parton energy loss} is not general enough to apply to the situations with an important interference between initial and final state radiation, where the associated {\it radiated energy} should instead be considered. Quite interestingly, when progressively going from large $\xf$ (or large rapidity) to the central rapidity region, we expect a transition between a regime where the (average) medium-induced radiated energy scales as $E$ and a regime where it does not. It would be interesting to single out some production process where such a transition can be put in evidence. In this respect, the RHIC data on quarkonium suppression in nuclear collisions appear very promising. The PHENIX collaboration reported a stronger $\jpsi$ suppression at forward rapidity than at mid-rapidity in $d$--Au~\cite{Adare:2007gn} and also Au--Au collisions~\cite{Adare:2006ns}, unlike the models based on quarkonium dissociation in QGP. Another observation, reported by STAR, is the smaller $\jpsi$ suppression when going from low to high $\pt$~\cite{Abelev:2009qaa}. Both measurements seem consistent with such a transition between the two identified regimes. More generally, we mentioned that some apparently unrelated nuclear suppression effects (jet-quenching in A--A collisions, open and hidden heavy-flavour and light-hadron production in $p$--A) might be explained in a unified framework. 

We also pointed out in the Introduction that the scaling $\Delta E \propto E$ predicted in the present paper for $\jpsi$ forward production would lead to the scaling of $\jpsi$ suppression with $x_1$ (or $\xf$) independently of the c.m. energy $\sqrt{s}$ of the $p$--A collision, 
as observed experimentally. On the contrary, an energy loss independent of the parton energy -- as for instance in Drell-Yan production -- would lead to a stronger  nuclear suppression at lower beam energy, since the typical shift of the momentum fraction in the proton PDF would scale like $\Delta x_1 \propto 1/E_{\rm beam} \propto 1/s$ causing a breakdown of $x_1$ (or $\xf$) scaling. Unlike the GM model we do not expect any scaling of DY suppression in either $x_1$ or $\xf$ variables. Lacking precise DY measurements at various beam energies, this prediction has not been tested yet. The future data to be collected by the E906 experiment ($E_{\rm beam}=120$~GeV)~\cite{e906} at Fermilab and possibly at J-PARC ($E_{\rm beam}=50$~GeV)~\cite{jparc} will allow for crucial tests of the energy loss effects discussed in this paper, both in DY and $\jpsi$ production. 

On a more theoretical ground, it would be valuable to study how our
results are related to QCD factorization issues. The physics described
in our study is clearly of collinear nature, as can be seen for
instance from the divergence of our result
\eq{indspec-quarkonium-final} (or its abelian version
\eq{indspeclargexf3-b}) in the formal limit $M_\perp \to 0$. However,
as we already mentioned, our analysis requires $M_\perp \gg k_\perp$,
$\Delta q_\perp$, so that the kinematics of the hard process is not
affected by (soft) radiation and rescatterings. As a consequence, the
average radiated energy is suppressed by a power of $M_\perp$ (see
\eq{largexfelectronloss} and \eq{quarkoniumloss}), \ie, it is of
higher-twist. Thus, our results should not conflict with leading-twist
collinear factorization~\cite{Collins:1989gx}. However, we do not see
why our effect should disappear when the kinematics is extended to the
domain where $M_\perp$, $\Delta q_\perp$ and $k_\perp$ are of the same
order. As already mentioned, in QED one may consider the incoming and outgoing energetic particles participating to the hard process to carry different charges $e$ and $e'$. The fact that the medium-induced radiation spectrum \eq{indspeclargexf3-b} is then proportional to $e e'$ suggests that the effect cannot be attributed to any purely initial (parton density) or final (fragmentation) effect. For the same reason, that the effect could be part of other (more general) universal objects, such as parton correlation functions (\ie, fully unintegrated parton densities)~\cite{Collins:2007ph}, seems also unlikely. It has been recently argued that factorization (even in a generalized, $k_T$-dependent formulation) is most probably violated in some cases, for instance in the production of large-$p_T$ back-to-back hadrons in hadron-hadron collisions~\cite{Collins:2007nk}. Similarly, it is plausible that hard-scattering factorization is truly violated in the small angle scattering process we have considered in the present study. 

%=========================================================
\vspace{0.4cm}
\begin{acknowledgments} 
\vspace{-0.4cm}
We thank S.~Brodsky and P.~Hoyer for valuable comments. FA would like to thank also CERN-TH for hospitality.
\end{acknowledgments}
%================================================

\providecommand{\href}[2]{#2}\begingroup\raggedright\endgroup

%==========================================================

\begin{thebibliography}{10}

\bibitem{Adler:2003qi}
{\bf PHENIX}, S.~S. Adler {\it et~al.},  Phys. Rev. Lett. {\bf 91} (2003)
  072301; 
%%CITATION = NUCL-EX 0304022;%%
{\bf STAR}, C.~Adler {\it et~al.},  Phys. Rev. Lett. {\bf 89} (2002) 202301.
%%CITATION = NUCL-EX 0206011;%%

\bibitem{Aamodt:2010jdCollaboration:2010bu}
{\bf ALICE}, K.~Aamodt {\it et~al.},
  \href{http://arXiv.org/abs/1012.1004}{{\tt 1012.1004}};\\
%%CITATION = 1012.1004;%%
{\bf ATLAS} Collaboration,  Phys. Rev. Lett. {\bf 105} (2010) 252303, \href{http://arXiv.org/abs/1011.6182}{{\tt 1011.6182}}.
%%CITATION = 1011.6182;%%

\bibitem{Accardi:2009qv}
A.~Accardi, F.~Arleo, W.~K. Brooks, D.~d'Enterria and V.~Muccifora, Riv.\ Nuovo\ Cim. {\bf 032} (2010) 439;
%%CITATION = 0907.3534;%%
D.~d'Enterria,   Landolt-Boernstein Vol. 1-23A (Springer Verlag), \href{http://arXiv.org/abs/0902.2011}{{\tt 0902.2011}}.
%%CITATION = 0902.2011;%%

\bibitem{Peigne:2008wu}
S.~Peign{{\'e}} and A.~V. Smilga,  Phys.\ Usp.\ {\bf 52} (2009) 659, \href{http://arXiv.org/abs/0810.5702}{{\tt
  0810.5702}}.
%%CITATION = 0810.5702;%%

\bibitem{Badier:1983dg}
{\bf NA3}, J.~Badier {\it et~al.},  Z. Phys. {\bf C20} (1983) 101.
%%CITATION = ZEPYA,C20,101;%%

\bibitem{Alde:1990wa}
{\bf E772}, D.~M. Alde {\it et~al.},  Phys. Rev. Lett. {\bf 66} (1991)
  133.
%%CITATION = PRLTA,66,133;%%

\bibitem{Kowitt:1993ns}
M.~S. Kowitt {\it et~al.},  Phys. Rev. Lett. {\bf 72} (1994) 1318.
%%CITATION = PRLTA,72,1318;%%

\bibitem{Leitch:1999ea}
{\bf E866}, M.~J. Leitch {\it et~al.},  Phys. Rev. Lett. {\bf 84} (2000)
  3256.
%%CITATION = NUCL-EX 9909007;%%

\bibitem{e866opencharm}
{\bf E866}, S.~A. Klinksiek, J.-C. Peng and P.~E. Reimer,  Workshop on
  ``High-energy Hadron Physics with Hadron Beams'', KEK (January 6-8, 2010).

\bibitem{Fischer:2002qp}
{\bf NA49}, H.~G. Fischer,  Nucl. Phys. {\bf A715} (2003) 118.
%%CITATION = HEP-EX/0209043;%%

\bibitem{Arsene:2003yk}
{\bf BRAHMS}, I.~Arsene {\it et~al.},  Phys. Rev. Lett. {\bf 91} (2003) 072305; {\it ibid},  {\bf 93} (2004) 242303.
%%CITATION = NUCL-EX/0307003;%%
%%CITATION = NUCL-EX/0403005;%%

\bibitem{Badier:1981ci}
{\bf NA3}, J.~Badier {\it et~al.},  Phys. Lett. {\bf B104} (1981) 335;
%%CITATION = PHLTA,B104,335;%%
{\bf E772}, D.~M. Alde {\it et~al.},  Phys. Rev. Lett. {\bf 64} (1990)
  2479;
{\bf E866}, M.~A. Vasilev {\it et~al.},  Phys. Rev. Lett. {\bf 83} (1999)
  2304.
%%CITATION = HEP-EX/9906010;%%

\bibitem{Adler:2005ph}
{\bf PHENIX}, S.~S. Adler {\it et~al.},  Phys. Rev. Lett. {\bf 96} (2006)
  012304.
%%CITATION = NUCL-EX 0507032;%%

\bibitem{Hoyer:1990us}
P.~Hoyer, M.~Vanttinen and U.~Sukhatme,  Phys. Lett. {\bf B246} (1990)
  217.
%%CITATION = PHLTA,B246,217;%%

\bibitem{Brodsky:1989ex}
S.~J. Brodsky and P.~Hoyer,  Phys. Rev. Lett. {\bf 63} (1989) 1566.
%%CITATION = PRLTA,63,1566;%%

\bibitem{Vogt:1999dw}
R.~Vogt,  Phys. Rev. {\bf C61} (2000) 035203.
%%CITATION = HEP-PH 9907317;%%

\bibitem{Lourenco:2008sk}
C.~Louren{\c c}o, R.~Vogt and H.~K. Woehri,  JHEP {\bf 02} (2009) 014.
%%CITATION = 0901.3054;%%

\bibitem{Arleo:1999af}
F.~Arleo, P.-B. Gossiaux, T.~Gousset and J.~Aichelin,  Phys. Rev. {\bf C61}
  (2000) 054906.
%%CITATION = HEP-PH/9907286;%%

\bibitem{Gavin:1991qk}
S.~Gavin and J.~Milana,  Phys. Rev. Lett. {\bf 68} (1992) 1834.
%%CITATION = PRLTA,68,1834;%%

\bibitem{Brodsky:1992nq}
S.~J. Brodsky and P.~Hoyer,  Phys. Lett. {\bf B298} (1993) 165.
%%CITATION = HEP-PH/9210262;%%

\bibitem{Baier:1996kr}
R.~Baier, Y.~L. Dokshitzer, A.~H. Mueller, S.~Peign{\'e} and D.~Schiff,  Nucl.
  Phys. {\bf B483} (1997) 291;
%%CITATION = HEP-PH/9607355;%%
B.~G. Zakharov,  JETP Lett. {\bf 65} (1997) 615.
%%CITATION = HEP-PH 9704255;%%

\bibitem{Kopeliovich:2005ym}
B.~Z. Kopeliovich, J.~Nemchik, I.~K. Potashnikova, M.~B. Johnson and
  I.~Schmidt,  Phys. Rev. {\bf C72} (2005) 054606;
%%CITATION = HEP-PH/0501260;%%
J.~Nemchik, V.~Petracek, I.~K. Potashnikova and M.~Sumbera,  Phys. Rev. {\bf
  C78} (2008) 025213.
%%CITATION = 0805.4267;%%

\bibitem{Weinberg:1965nx}
S.~Weinberg,  Phys. Rev. {\bf 140} (1965) B516.
%%CITATION = PHRVA,140,B516;%%

\bibitem{Dokshitzer:1977sg}
Y.~L. Dokshitzer,  Sov. Phys. JETP {\bf 46} (1977) 641;
%%CITATION = SPHJA,46,641;%%
V.~N. Gribov and L.~N. Lipatov,  Sov. J. Nucl. Phys. {\bf 15} (1972) 438;
%%CITATION = SJNCA,15,438;%%
L.~N. Lipatov,  Sov. J. Nucl. Phys. {\bf 20} (1975) 94;
%%CITATION = SJNCA,20,94;%%
G.~Altarelli and G.~Parisi,  Nucl. Phys. {\bf B126} (1977) 298.
%%CITATION = NUPHA,B126,298;%%

\bibitem{Bodwin:1994jh}
G.~T. Bodwin, E.~Braaten and G.~P. Lepage,  Phys. Rev. {\bf D51} (1995)
  1125.
%%CITATION = HEP-PH/9407339;%%

\bibitem{Fritzsch:1977ay}
H.~Fritzsch,  Phys. Lett. {\bf B67} (1977) 217;
%%CITATION = PHLTA,B67,217;%%
F.~Halzen,  Phys. Lett. {\bf B69} (1977) 105;
%%CITATION = PHLTA,B69,105;%%
J.~F. Amundson, O.~J.~P. {\'E}boli, E.~M. Gregores and F.~Halzen,  Phys. Lett.
  {\bf B390} (1997) 323.
%%CITATION = HEP-PH 9605295;%%

\bibitem{Chang:1979nn}
C.-H. Chang,  Nucl. Phys. {\bf B172} (1980) 425;
%%CITATION = NUPHA,B172,425;%%
R.~Baier and R.~R{\"u}ckl,  Phys. Lett. {\bf B102} (1981) 364;
%%CITATION = PHLTA,B102,364;%%
{\it ibid.},  Z. Phys. {\bf C19} (1983) 251.
%%CITATION = ZEPYA,C19,251;%%

\bibitem{Hoyer:1997yf}
P.~Hoyer and S.~Peign{\'e},  Phys. Rev. {\bf D57} (1998) 1864.
%%CITATION = HEP-PH/9706486;%%

\bibitem{Arleo:2003jz}
F.~Arleo,  Eur. Phys. J. {\bf C30} (2003) 213;
%%CITATION = HEP-PH 0306235;%%
A.~Accardi,  Phys. Rev. {\bf C76} (2007) 034902.
%%CITATION = 0706.3227;%%

\bibitem{Lansberg:2008gk}
J.-P. Lansberg,  Eur. Phys. J. {\bf C61} (2009) 693.
%%CITATION = 0811.4005;%%

\bibitem{Vogt:2001ky}
R.~Vogt,  Nucl. Phys. {\bf A700} (2002) 539.
%%CITATION = HEP-PH 0107045;%%

\bibitem{Adare:2007gn}
{\bf PHENIX}, A.~Adare {\it et~al.},  Phys. Rev. {\bf C77} (2008) 024912.
%%CITATION = 0711.3917;%%

\bibitem{Adare:2006ns}
{\bf PHENIX}, A.~Adare,  \href{http://arXiv.org/abs/nucl-ex/0611020}{{\tt
  nucl-ex/0611020}}.
%%CITATION = NUCL-EX 0611020;%%

\bibitem{Abelev:2009qaa}
{\bf STAR}, B.~I. Abelev {\it et~al.},  Phys. Rev. {\bf C80} (2009) 041902.
%%CITATION = 0904.0439;%%

\bibitem{e906}
{\bf E906}, D. Geesaman,
P. Reimer, {\em et al.}, Fermilab E906 (1999), \url{http://www.phy.anl.gov/mep/drell-yan}

\bibitem{jparc}
Workshop on High-energy hadron physics with hadron beams, KEK Tsukuba (Japan), 6-8 January 2010, \url{http://www-conf.kek.jp/hadron1/hehp-th10}

\bibitem{Collins:1989gx}
J.~C. Collins, D.~E. Soper and G.~Sterman,  Adv. Ser. Direct. High Energy Phys.
  {\bf 5} (1988) 1.
%%CITATION = HEP-PH/0409313;%%

\bibitem{Collins:2007ph}
J.~C. Collins, T.~C. Rogers and A.~M. Sta\'sto,  Phys. Rev. {\bf D77} (2008)
  085009.
%%CITATION = 0708.2833;%%

\bibitem{Collins:2007nk}
J.~Collins and J. Qiu,  Phys. Rev. {\bf D75} (2007) 114014.
%%CITATION = 0705.2141;%%

\bibitem{Vitev:2007ve}
I.~Vitev,  Phys. Rev. {\bf C75} (2007) 064906.
%%CITATION = HEP-PH/0703002;%%

\end{thebibliography}
\end{document}